\documentclass[preprint]{aastex}
\usepackage{color, soul}
\usepackage{graphicx}
\usepackage{subfigure}
\usepackage{rotating}
\usepackage{amsmath}

\begin{document}
\title{{\it Chandra} Detection of Intra-cluster X-ray sources in Virgo}

\author{Meicun Hou\altaffilmark{1,2}, Zhiyuan Li\altaffilmark{1,2}, Eric W. Peng\altaffilmark{3,4}, Chengze Liu\altaffilmark{5,6}}
\affil{$^{1}$ School of Astronomy and Space Science, Nanjing University, Nanjing 210023, China}
\affil{$^{2}$ Key Laboratory of Modern Astronomy and Astrophysics (Nanjing University), Ministry of Education, Nanjing 210023, China}
\affil{$^{3}$ Department of Astronomy, Peking University, Beijing 100871, China}
\affil{$^{4}$ Kavli Institute for Astronomy and Astrophysics, Peking University, Beijing 100871, China}
\affil{$^{5}$ Center for Astronomy and Astrophysics, Department of Physics and Astronomy, Shanghai Jiao Tong University, Shanghai 200240, China}
\affil{$^{6}$ Shanghai Key Lab for Particle Physics and Cosmology, Shanghai Jiao Tong University, Shanghai 200240, China}
\email{lizy@nju.edu.cn}

\begin{abstract}
We present a survey of X-ray point sources in the nearest and dynamically young galaxy cluster, Virgo, using archival {\it Chandra} observations that sample the vicinity of 80 early-type member galaxies. 
The X-ray source populations at the outskirt of these galaxies are of particular interest. 
We detect a total of 1046 point sources (excluding galactic nuclei) out to a projected galactocentric radius of $\sim$40 kpc and down to a limiting 0.5-8 keV luminosity of $\sim$$2\times10^{38}{\rm~erg~s^{-1}}$. 
Based on the cumulative spatial and flux distributions of these sources, we statistically identify $\sim$120 excess sources that are not associated with the main stellar content of the individual galaxies, nor with the cosmic X-ray background.
This excess is significant at a 3.5\,$\sigma$ level, when Poisson error and cosmic variance are taken into account. 
On the other hand, no significant excess sources are found at the outskirt of a control sample of field galaxies, suggesting that at least some fraction of the excess sources around the Virgo galaxies are truly intra-cluster X-ray sources.
Assisted with ground-based and HST optical imaging of Virgo, we discuss the origins of these intra-cluster X-ray sources, in terms of supernova-kicked low-mass X-ray binaries (LMXBs), globular clusters, LMXBs associated with the diffuse intra-cluster light, stripped nucleated dwarf galaxies and free-floating massive black holes.
\end{abstract}


\section{Introduction} \label{sec:intro}

X-ray studies of nearby normal galaxies have established two primary components of the X-ray emission: point sources and diffuse hot gas (e.g., \citealp{tri85,2000ApJ...544L.101S}).
The point sources, typically detected with luminosities $L_{\rm X} \gtrsim10^{37}{\rm~erg~s^{-1}}$, consist of high- and low-mass X-ray binaries (HMXBs and LMXBs).   
Due to their long evolution timescales, LMXBs are often considered a good tracer of old stellar populations. In particular, it has been shown that both the total number and cumulative luminosity of LMXBs in early-type galaxies (ETGs) are a robust, quasi-linear indicator of the host galaxy's stellar mass (\citealp{2004MNRAS.349..146G}). 
This empirical relation, however, is complicated by the contribution of LMXBs residing in globular clusters (hereafter called GC-LMXBs). 
Generally thought to be the product of stellar encounters within the dense GC environment, GC-LMXBs of a given galaxy should exist in a disproportionate number with respect to the total stellar mass. 
Isolating the GC-LMXBs requires high-resolution and high-sensitivity optical/X-ray observations, which are currently available for only a small number of nearby ETGs (e.g., Kim et al.~2009, 2013; Zhang et al.~2011; Lehmer et al.~2014; Peacock \& Zepf 2016).

A further complication emerges when the spatial distribution of X-ray sources in and around ETGs is carefully examined. Using deep {\it Chandra} observations, Li et al.~(2010) identified $\sim$100 X-ray sources between 2-5 effective radii of the Sombrero galaxy (M\,104), and found that the cosmic X-ray background (CXB) can account for only half of them, suggesting that the rest ($\sim$50) be physically related to the halo of M\,104, with $L_{\rm X} \approx 10^{37-38}{\rm~erg~s^{-1}}$.
This was a rather surprising result, since the bulk of the stellar content in M\,104 should lie within its bulge and disk. 
Zhang, Gilfanov \& Bogd{\'a}n (2013) further investigated this issue, finding a similar excess of halo X-ray sources in a sizable sample of nearby ETGs.
The significance of the excess was apparently correlated with both the so-called GC specific frequency and the stellar mass of the host galaxy. 
This led the authors to propose that the halo excess mainly consists of (i) GC-LMXBs, which tend to have a broader radial distribution than the stellar spheroid, and (ii) neutron star-LMXBs having received a strong supernova kick and subsequently escaped from their birth place into the halo.

Notably, many of the ETGs studied in Zhang et al.~(2013), as well as M\,104, are relatively isolated galaxies. It remains to be determined whether environmental effects play a role in the formation of an extended halo of X-ray sources around ETGs.   
Inside the dense environment of clusters and groups, galaxy-galaxy interactions, such as tidal encounters and mergers, continue to shape and redistribute the stellar content of essentially all member galaxies. 
Consequently, extended envelopes of stars grow in the most massive galaxies, whereas small galaxies are stripped off of their stars, which either directly merge into massive galaxies or spread over the intergalactic space. 
These processes, generic in galaxy clusters, lead to the gradual building of the so-called diffuse intra-cluster light (ICL; \citealp{mih15} and references therein).    
Direct probe of the ICL in the optical proves to be challenging due to their low surface brightness.  
Nevertheless, various studies have revealed the presence of ICL in galaxy clusters from local to intermediate redshifts ($z \sim 0.5$), which typically amount to $\sim$10-30\% of the cluster's total stellar mass (e.g., \citealp{mih05, zib05, bur15, mor16}).  
In particular, the ICL in Virgo, the nearest and dynamically young cluster, has been studied in detail and estimated to account for a fractional stellar mass of 7-15\% (\citealp{mih05,mih17}).

It is conceivable that the ICL, consisting of predominantly old stellar populations, also harbor LMXBs, the abundance of which, to first order, should follow the aforementioned quasi-linear relation with stellar mass. Conversely, the existence of such LMXBs can potentially offer a new, effective way to probe the ICL, at least in the local Universe where X-ray observations are capable of resolving LMXBs. 
Moreover, during the course of the hierarchical assembly of galaxy clusters, exotic objects other than LMXBs might have been released into the intra-cluster space. 
In particular, a (super-)massive black hole ((S)MBH) can be ejected from its host galaxy due to strong gravitational recoil following the merger of two progenitor (S)MBHs (see \citealp{sch13} for a review). 
In addition, a MBH might be exposed after the near-complete stripping of its parent stellar spheroid in the cluster tidal field. 
In both cases, the nearly bare MBHs wondering in the cluster may manifest itself as an X-ray emitter, powered by low-level accretion due to the paucity of surrounding medium (e.g., \citealp{mer09}), yet detectable with sensitive X-ray observations.

In recent years, with the advent of {\it the Chandra Observatory}, there have been extensive studies of the X-ray behavior of galaxies in group/cluster environments, e.g., hot gas halo (Jeltema et al.~2007,2008; Desjardins et al.~2014), AGN content (Shen et al.~2007; Miller et al.~2012a; Tzanavaris et al.~2014), X-ray binary-galaxy scaling relations (Finoguenov et al.~2004; Hornschemeier et al. 2006; Tzanavaris et al.~2016), among others. 
However, a detailed study of the cluster X-ray source populations in a spatially-resolved fashion is still absent. 

We are thus motivated to carry out the present work. Based on archival {\it Chandra} observations, we search for X-ray sources in and around a representative sample of ETGs in Virgo, and provide strong evidence for the existence of intra-cluster X-ray source populations. 
Our sample selection and source detection procedure are described in Section~\ref{sec:data}. 
Analysis of the source properties, in particular their spatial distribution with respect to the member galaxies, is presented in Section~\ref{sec:analysis}. The implications of our findings are discussed in Section~\ref{sec:discussion}, followed by a summary in Section \ref{sec:sum}. 
Throughout this work, we adopt a uniform distance of 16.5 Mpc (1$^{\prime\prime}$ corresponds to 80 pc; \citealp{mei07}) for all sources and galaxies in Virgo. The line-of-sight depth of Virgo introduces an uncertainty of no more than ten percent in the derived luminosities, which should have little effect in our results. 
Errors are quoted at 1\,$\sigma$ confidence level, unless otherwise stated.

\section{Data Preparation} \label{sec:data}
\subsection{Sample selection} \label{subsec:sample}
\begin{figure*}\centering
\includegraphics[width=0.9\textwidth,angle=0]{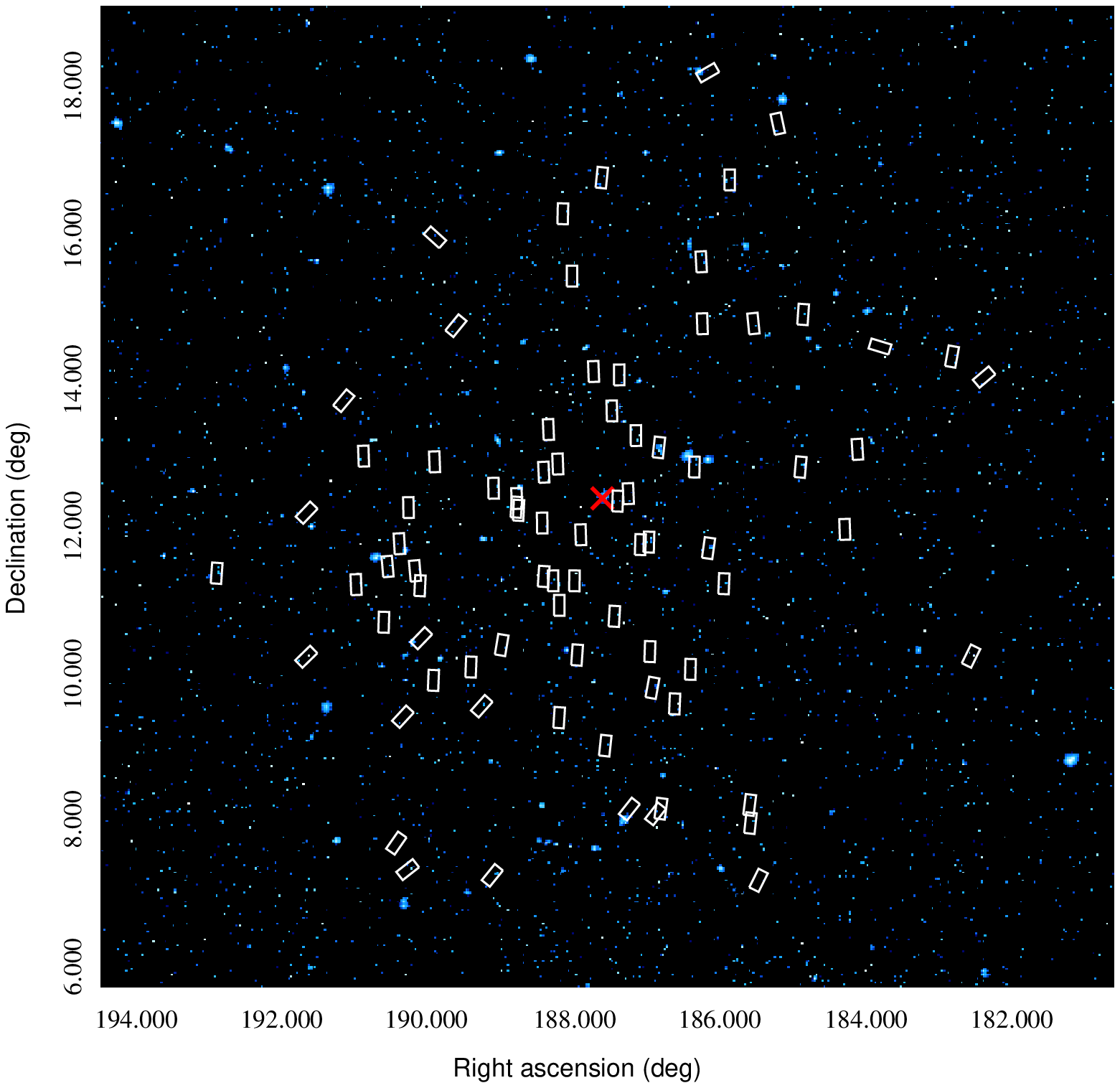}
\caption{A Sloan Digital Sky Survey $r$-band image of the Virgo cluster. The 80 AMUSE-Virgo {\it Chandra}/ACIS fields are outlined by the white boxes. The center of M\,87 is marked by a red cross.
}
\label{fig:fov}
\end{figure*}

The Virgo cluster spans $\sim$100 deg$^2$ on the sky, making any survey of its entirety a challenging task. To date, a full mapping of Virgo has only been conducted by {\it ROSAT} (B\"{o}hringer et al.~1994), which will be superceded by the upcoming eROSITA mission (Merloni et al.~2012).
Nevertheless, a large number of {\it Chandra} observations exist that are pointed towards individual member galaxies, providing the best angular resolution for identifying point sources.
For our purpose, we originally selected 84 early-type member galaxies, including ellipticals, lenticulars and dwarf spheroidals, which are the prime targets of the AGN Multiwavelength Survey of Early-Type Galaxies in the Virgo Cluster (AMUSE-Virgo, \citealp{2008ApJ...680..154G,2010ApJ...714...25G}), a {\it Chandra} Large Program that aimed to provide an unbiased census of SMBH activity in the cluster environment.
Each of the AMUSE-Virgo galaxies has a snapshot exposure of $\sim$5 ks with the Advanced CCD Imaging Spectrometer (ACIS), ensuring a quasi-uniform, albeit moderate, sensitivity for source detection of $\sim$$1.8\times10^{38}{\rm~erg~s^{-1}}$ in 0.5-8 keV band, which is crucial for characterizing the source spatial distribution (Section~\ref{sec:analysis}).   
A further advantage of the AMUSE-Virgo fields lies in that they provide a good sampling of the cluster volume (Figure~\ref{fig:fov}).  
The large number of galaxies involved also holds promise for a meaningful statistical study of the X-ray source populations.

Among the 84 fields, two are located within a projected distance of $10^{\prime}$ ($\sim$50 kpc) from the center of M\,87, hence any source to be found there might be physically associated with the extended, X-ray-bright halo of M\,87. 
We excluded these two fields to avoid confusion with the potentially complex stellar populations of M\,87, as well as with small-scale diffuse features, or ``clumps'', arising from the hot gas (Forman et al.~2007). A dedicated study of the X-ray sources around M\,87 will be presented elsewhere (L. Luan et al. in preparation).  
For similar consideration we also excluded two fields that are located within $5^{\prime}$ from the center of the giant elliptical NGC\,4472.
Table~\ref{tab:log} presents the basic information, including galaxy name, position, ObsID, and effective exposure, of the remaining 80 AMUSE-Virgo fields.

To facilitate a direct comparison of the X-ray source distribution in different galactic environments, it is desired to have a control sample of galaxies not residing in clusters. 
We find that such a control sample is readily offered by the AMUSE-Field program \citep{2012ApJ...747...57M}, which surveyed the SMBH activity in a distance-limited ($\lesssim$ 30 Mpc) sample of field ETGs, in parallel to the AMUSE-Virgo sample. 
Therefore, we selected 57 AMUSE-Field galaxies, each of which has an ACIS exposure ranging between 2-12 ks. 
By design, this ensures a quasi-uniform limiting luminosity of $\sim$$2\times10^{38}{\rm~erg~s^{-1}}$, similar to that for the AMUSE-Virgo galaxies. 

\subsection{X-ray source detection} \label{subsec:detection}

For all the selected ACIS fields, we downloaded and reprocessed the archival data using CIAO v4.8 and the corresponding calibration files, following the standard procedure\footnote{http://cxc.harvard.edu/ciao/}. 
For each observation, we produced counts and exposure maps on the original pixel scale ($0 \farcs 492 {\rm~pixel^{-1}}$) in the 0.5-2 ($S$), 2-8 ($H$), and 0.5-8 ($F$) keV bands. 
The exposure maps were weighted by a fiducial incident spectrum, which is an absorbed power-law with a photon-index of 1.7 and an absorption column density $N_{\rm H}$ = $10^{21} {\rm~cm^{-2}}$. 
The latter value is somewhat higher than the Galactic foreground absorption column of $\sim$$2.5\times10^{20} {\rm~cm^{-2}}$ towards Virgo, but allows for some internal absorption. 
We examined the light curve of each observation and found that the instrumental background was quiescent in the vast majority of cases. 
Hence we decided to preserve all the science exposure so that the detection sensitivity remains highly uniform among all the fields. 

\begin{figure*}\centering
\includegraphics[width=0.9\textwidth, angle=0]{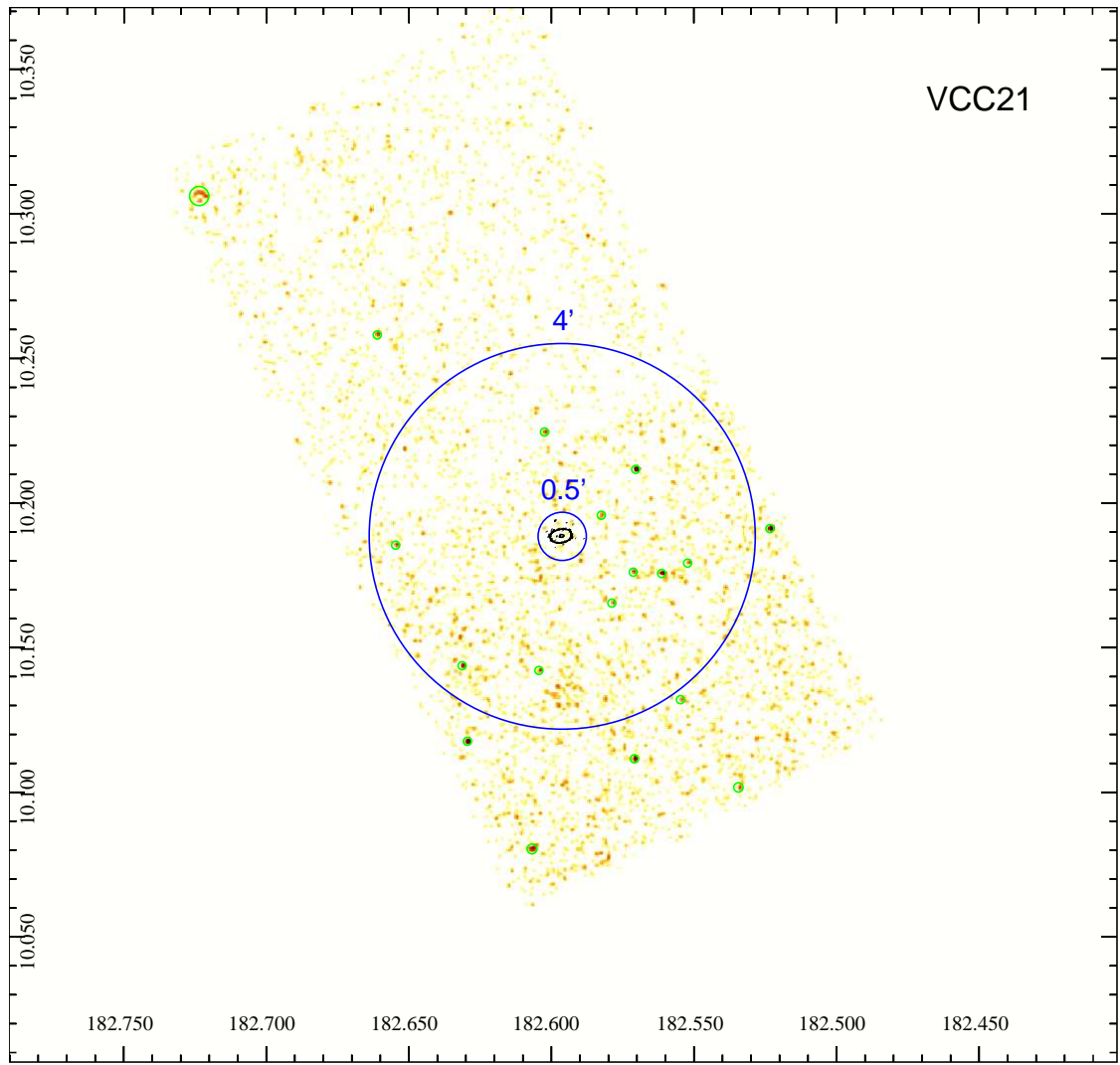}
\caption{0.5-8 keV counts image of galaxy VCC21. The detected X-ray sources are marked by green circles (with different sizes representing the 90\% EER). 
The optical extent of the galaxy is outlined by the black contours, which are derived from the ACSVCS survey.
The small blue circle (with a radius of 0\farcm5) illustrates the region within which the majority of the galaxy's LMXBs are expected to be found,
while the large blue circle delineates the radius ($4^{\prime}$) up to which a full azimuthal coverage is achieved.  
}
\label{fig:vcc21}
\end{figure*}

Following the source detection procedure detailed in Wang~(2004) and Li et al.~(2010), 
we detected X-ray sources in the $S$, $H$ and $F$ bands for each field.
The procedure employed a combination of source detection algorithms: wavelet, sliding-box and maximum likelihood centroid fitting. 
In all cases, the target galaxy was placed at the aim-point on the S3 CCD.
Only data from the S3 and S2 CCDs were used, to ensure optimal sensitivity for source detection.
We note that this choice allows for full azimuthal coverage up to a radius of $4^\prime$ from the center of each target galaxy.
The field of galaxy VCC\,21 is shown as an example in Figure~\ref{fig:vcc21}.
The 80 AMUSE-Virgo fields together cover 3.1 deg$^2$ of the Virgo cluster,
from which we detected a total of 1234 sources, with a local false detection probability $P \leq {10^{-6}}$ (empirically yielding $\sim$0.1 false detection per field; Wang 2004).
The global distribution of the detected sources in Virgo is shown in Figure 1.

For each detected source, background-subtracted and exposure map-corrected count rates in the individual bands are derived from within the 90\% enclosed-energy radius (EER), taking into account the position-dependence of the PSF and the local background. 
Towards large radii (off-axis angles), the PSF becomes increasingly larger, substantially degrading the detection sensitivity. 
There is also chance that small clumps of the intra-cluster medium (ICM) mimic the appearance of genuine point sources with a large PSF. However, we show below (Section~\ref{subsec:spectral}) that the cumulative spectrum of the detected sources is inconsistent with a thermal plasma, indicating that contamination of the ICM clumps is negligible.
According to the assumed incident power-law spectrum, we have adopted a $F$-band count rate-to-intrinsic flux conversion factor of $6.8\times10^{-12}{\rm~erg~s^{-1}~cm^{-2}/(\rm counts~s^{-1})}$.
For sources located at the distance of Virgo, this translates to a count rate-to-luminosity conversion factor of $2.2\times10^{41}{\rm~erg~s^{-1}/(\rm counts~s^{-1})}$.
The $F$-band detection limit for each field is given in Table 1. 
It is noteworthy that essentially in all fields the limiting luminosity just exceeds the Eddington limit for an accreting neutron star of 1.4 solar masses, i.e., $\sim$$1.8\times10^{38}{\rm~erg~s^{-1}}$.

We then filtered our raw source list for further analysis. 
First, we noticed that in seven occasions, two AMUSE-Virgo fields partially overlap with each other, in which a source may be repeatedly detected.
If the angular separation between two sources is smaller than the quadratic sum of their position errors, we considered them as the same source and counted only the one with the smaller position error. 
In this way, we removed 25 duplicate sources. 

Since we are not interested in the X-ray-emitting galactic nuclei, we further excluded any source
that is located within $3^{\prime\prime}$ from the optical center of the target galaxy. 
In this way, we found 22 nuclear sources, which is consistent with Gallo et al.~(2010). 
We also searched the extended Virgo Cluster catalog (EVCC, \citealp{2014ApJS..215...22K}) for any additional galaxies that fall within each field and are brighter than 16 mag in the $r$-band, which is roughly the limiting magnitude of the AMUSE-Virgo galaxies. 
We identified 4 fields with two additional EVCC galaxies and 19 fields with one additional EVCC galaxy. 
The additional EVCC galaxies are noted in Table 1; for the four fields with two additional galaxies, only the brighter one is taken into account.
Among these EVCC galaxies, we found only one (VCC\,1253) harboring an X-ray source at it nucleus, which is subsequently excluded.
In summary, we preserved 1186 sources for further analysis, among which 1046 were detected in the $F$-band, 916 in the $S$-band, and 337 in the $H$-band. 
The number of X-ray point sources in each field is listed in Table 1, along with the expected number of CXB sources (Section~\ref{subsec:properties}).

We applied the same detection procedure for the 57 AMUSE-Field observations and detected 1054 sources in total. 
After removing 4 duplicate detections and 18 sources found within $3^{\prime\prime}$ from the galactic center, we selected 1032 independent (909 sources in the $F$-band), X-ray point sources in and around the field galaxies as our control sample. 

\section{Analysis and results} \label{sec:analysis}
\subsection{X-ray flux distribution} \label{subsec:properties}
\begin{figure*}\centering
\includegraphics[width=0.9\textwidth,angle=90]{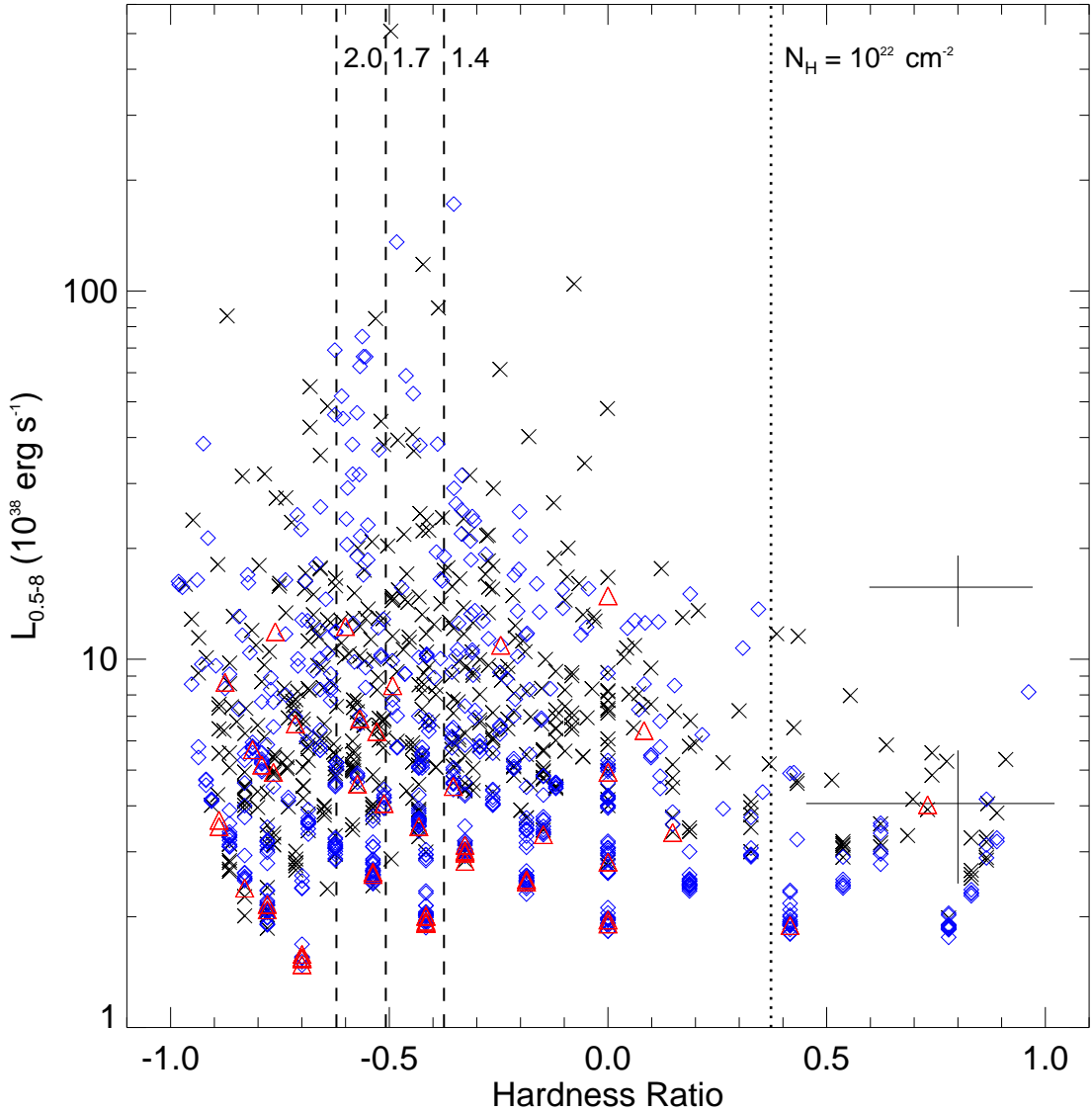}
\caption{0.5-8 keV luminosity vs. hardness ratio, defined as HR = $(H-S)/(H+S)$. The red triangles, blue diamonds and black crosses represent sources detected within $R < 0\farcm5$, $0\farcm5 < R < 4^{\prime}$, and $R > 4^{\prime}$, respectively, where $R$ is the projected radius from the center of the target galaxy. 
The vertical dashed lines, from left to right, correspond to hardness ratios for an absorbed power-law spectrum with a fixed $N_{\rm H}$ = $10^{21} {\rm~cm^{-2}}$ and a photon index of 2.0, 1.7, and 1.4, respectively. The vertical dotted lines correspond to hardness ratios for an absorbed power-law spectrum with a fixed photon index of 1.7 and $10^{22} {\rm~cm^{-2}}$.
The upper (lower) error bar on the right illustrates the median uncertainty in the hardness ratio and luminosity of sources more (less) luminous than $10^{39}{\rm~erg~s^{-1}}$.
}
\label{fig:HR}
\end{figure*}

We show the 0.5-8 keV intrinsic luminosity versus hardness ratio for the AMUSE-Virgo sources in Figure~\ref{fig:HR}. 
The source luminosity is calculated assuming a Virgo distance, although the majority of these sources should lie in the background (see below).
The hardness ratio, defined as $HR$ = $(H-S)/(H+S)$, is calculated with the observed counts in $S$ and $H$ bands using a Bayesian approach (\citealp{2006ApJ...652..610P}).
Also plotted for comparison are the predicted hardness ratios of certain absorbed power-law spectra, with varied absorption column and photon-index. 
It can be seen that most sources exhibit a hardness ratio consistent with being X-ray binaries or background AGN, which have a typical power-law spectrum with a photon-index of $1.4-2$.  
We have distinguished sources detected within three radial ranges: $R < 0\farcm5$ (red triangles), $0\farcm5 < R < 4^{\prime}$ (blue diamonds), and $R > 4^{\prime}$ (black crosses), where $R$ is the projected radius from the center of the target galaxy. 
Several sources found at $R > 0\farcm5$ have an apparent luminosity greater than $10^{40}{\rm~erg~s^{-1}}$, but these are most likely background sources.

\begin{figure*}\centering
\includegraphics[width=\textwidth,angle=90]{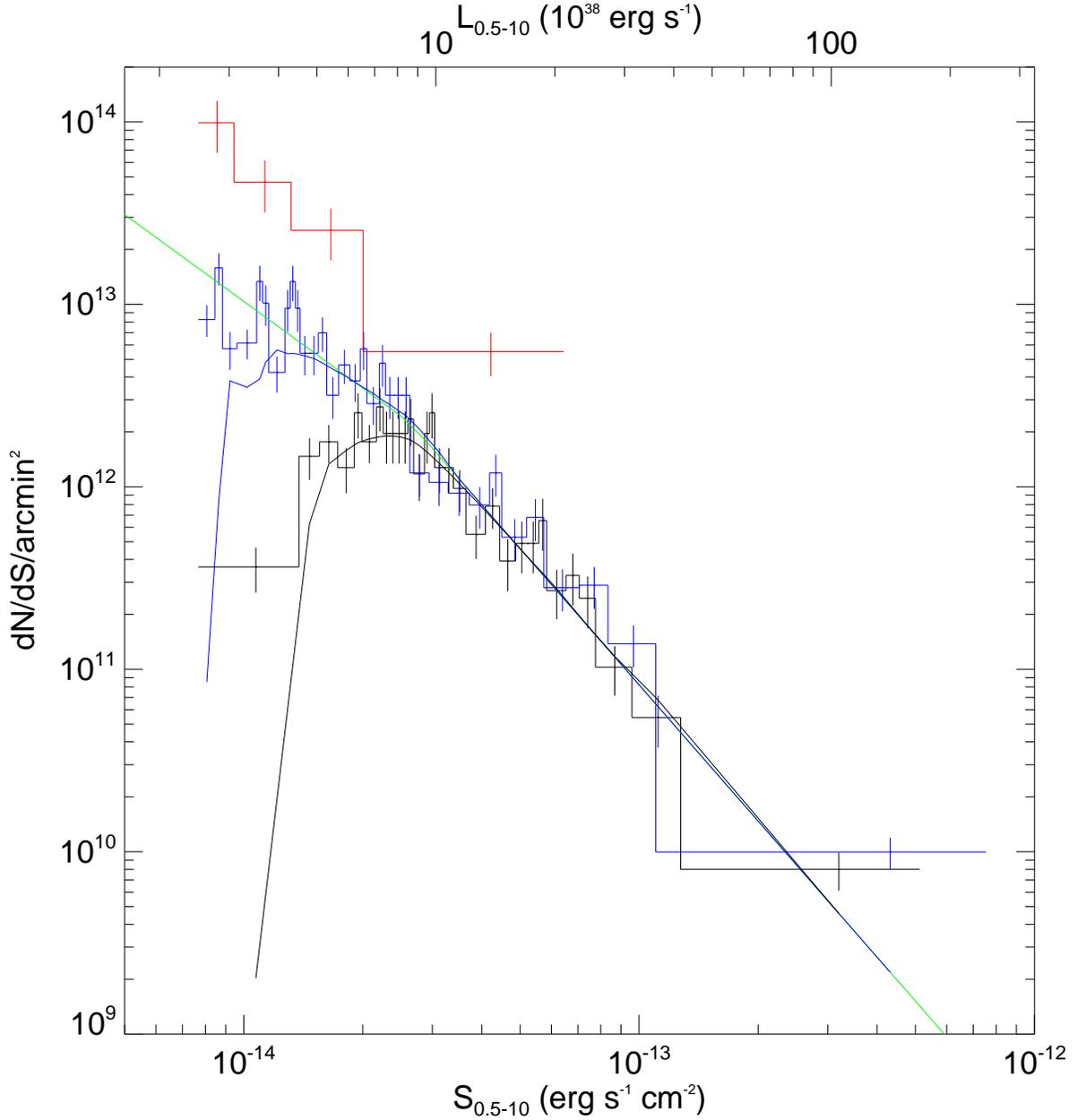}
\caption{The log$N$-log$S$ relation of sources detected in $B$-band and within $R < 8^\prime$. 
The source fluxes have been converted to 0.5-10 keV band for ease of comparison with the empirical log$N$-log$S$ relation of the cosmic X-ray background (green curve). The red, blue and black histograms represent sources falling within the three radial ranges as defined in Figure~\ref{fig:HR}. 
The black and blue curves are the corresponding cosmic X-ray background, corrected for detection incompleteness and Eddington bias at different fluxes. 
}
\label{fig:logN}
\end{figure*}

We further show in Figure~\ref{fig:logN} the differential flux density distribution (i.e., log$N$-log$S$ relation) of the 1007 sources detected in $F$-band and within $8^{\prime}$ from the center of the target galaxy, averaged over the 80 AMUSE-Virgo fields. 
While sources have been detected up to a projected radius of $10\farcm5$ in some fields, here we include only those within $R < 8^{\prime}$, such that the fractional azimuthal coverage at a given radius is at least $\sim$20\% (100\% within $R < 4^{\prime}$; Figure~\ref{fig:vcc21}). 
Similarly, we have distinguished sources falling within the three radial ranges (red for $R < 0\farcm5$, blue for $0\farcm5 < R < 4^{\prime}$ and black for $4^\prime < R < 8^{\prime}$).
 
The observed flux density distributions are to be contrasted with the empirical log$N$-log$S$ relation of the CXB, for which we adopt the broken power-law form of Georgakakis et al.~(2008),
\begin{equation}
\frac{dN}{df_{\rm x}} = 
\begin{cases}
K(\frac{f_{\rm x}}{f_{\rm ref}})^ {\beta_{\rm 1}},  & f_{\rm x} < f_{\rm b}\\
K'(\frac{f_{\rm x}}{f_{\rm ref}})^ {\beta_{\rm 2}}, & f_{\rm x} \geq f_{\rm b}
\end{cases}
\end{equation}
where $f_{\rm b}$ is the break flux and $f_{\rm ref}=10^{-14}\rm~erg~s^{-1}~cm^{-2}$; 
the normalizations $K$ and $K'$ are related to each other via $K'=K(f_{\rm b}/f_{\rm ref})^{\beta_{\rm 1}-\beta_{\rm 2}}$. 
The parameters for the CXB in the 0.5-10 keV band are $\beta_{\rm 1} =-1.58$, $\beta_{\rm 2} =-2.48$, $f_{\rm b}=2.63\times10^{-14}\rm~erg~s^{-1}~cm^{-2}$ and $K=3.74\times10^{16}\rm~deg^{-2}/(\rm erg~s^{-1}~cm^{-2})$.
We have converted the $F$-band count rate into 0.5-10 keV flux by assuming an intrinsic power-law spectrum with a photon-index of 1.4, suitable for the CXB. 
This empirical log$N$-log$S$ is plotted as a green curve in Figure~\ref{fig:logN}.

The observed number of sources is subject to detection incompleteness and the so-called Eddington bias. 
The former is due to variation in the effective exposure, local background and PSF across the field-of view, whereas the latter results from Poisson fluctuation in the detected count rate, coupled to the slope of the intrinsic log$N$-log$S$ relation. 
These two factors conspire to produce the apparent drop at the faint end in Figure~\ref{fig:logN}.
Therefore, we correct the empirical log$N$-log$S$ relation of the CXB for detection incompleteness and Eddington bias, 
and plot the results in Figure~\ref{fig:logN} (blue curve for $0\farcm5 < R < 4^{\prime}$ and black curve for $4^{\prime} < R < 8^{\prime}$). 
It can be seen that the sources detected within $4^{\prime} < R < 8^{\prime}$ agree quite well with the CXB. 
At $0\farcm5 < R < 4^{\prime}$, the detected sources are also dominated by the CXB, but excess to the CXB becomes evident.
On the other hand, the CXB has only a minor contribution at $R < 0\farcm5$, indicating that most sources detected there are associated with the target galaxies. 
In particular, the four sources with $L_{\rm X} \approx (1-2)\times10^{39}{\rm~erg~s^{-1}}$ are good candidates of black hole binaries.

\subsection{Spatial distribution of the X-ray sources} \label{subsec:spatial}
\begin{figure*}\centering
\includegraphics[scale=0.49, angle=90]{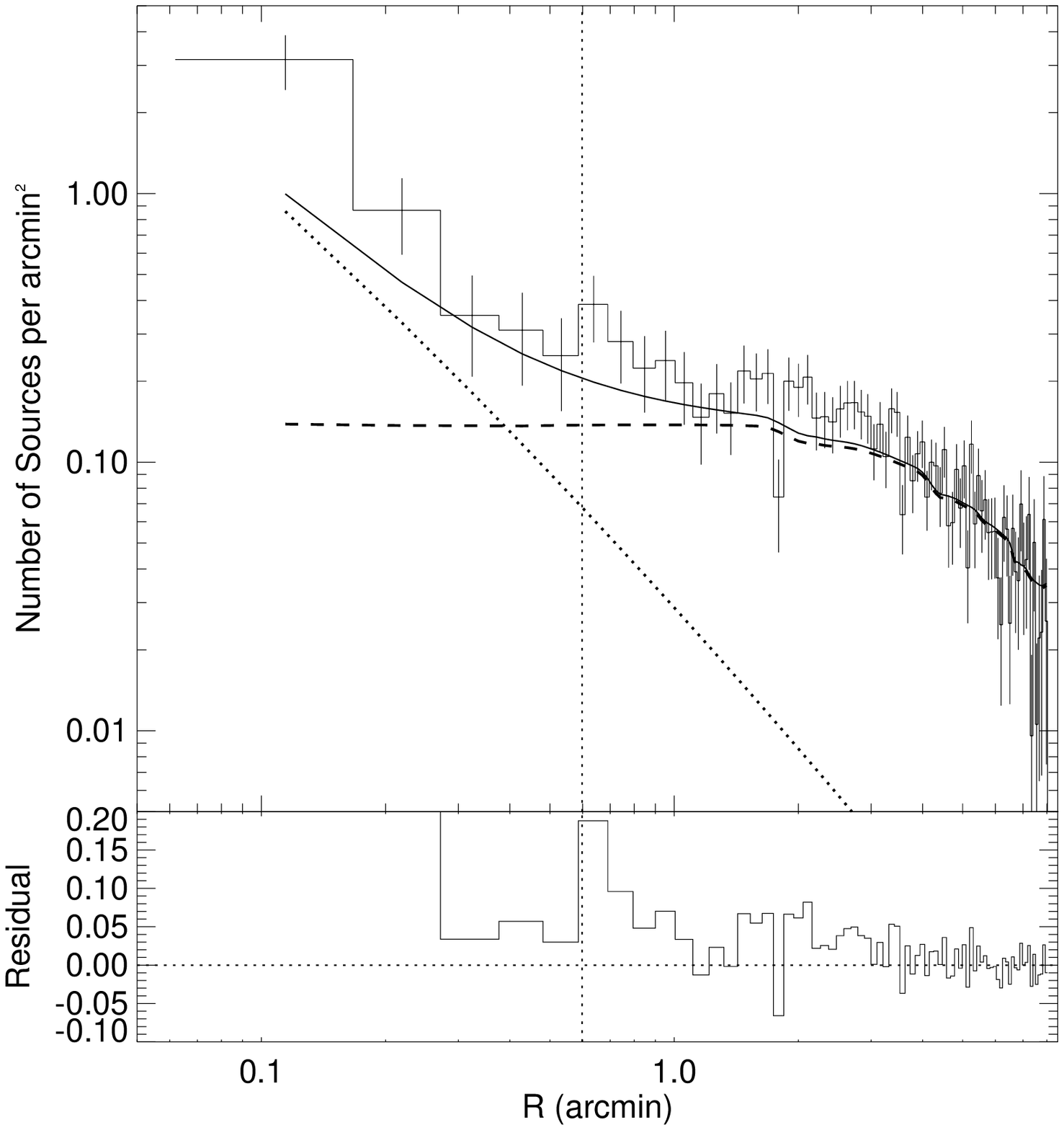}
\includegraphics[scale=0.49, angle=90]{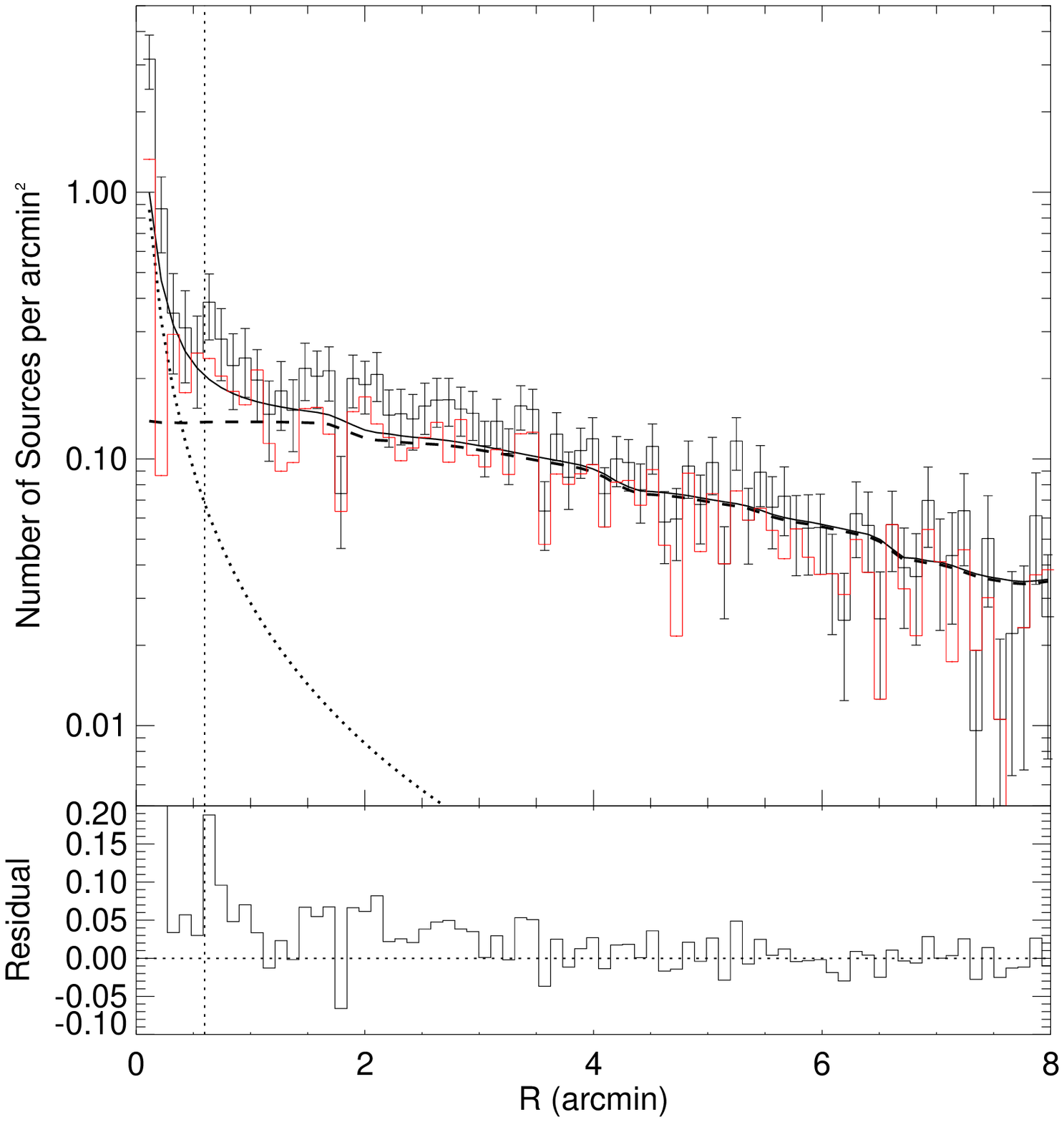}
\caption{{\it Left panel}: Radial surface density distribution of the $F$-band sources, averaged over the AMUSE-Virgo galaxies. 
The black histograms have the same width per bin. 
The dotted curve represents the estimated contribution of field-LMXBs associated with the target galaxies. 
The dashed curve represents the CXB, corrected for detection incompleteness and Eddington bias. 
The vertical dashed line marks the position of $R=3R_{\rm e}$, where $R_{\rm e} = 12\farcs0$ is the median effective radius of 80 target galaxies.
See text for details.
The lower panel shows the residual after subtracting the combined contribution (solid curve) of field-LMXBs and CXB, to highlight the excess within $4^\prime$.
{\it Right panel}: same as in the left panel, but with the x-axis in linear scale. The red histogram indicates the X-ray sources with an optical counterpart.
}
\label{fig:radial}
\end{figure*}

The spatial distribution of the X-ray sources offers further insight about their nature.
Figure~\ref{fig:radial} displays the $F$-band source surface density profile as a function of projected radius $R$, averaged over the 80 AMUSE-Virgo fields. 
The X-ray sources should consist of at least two main components: (i) X-ray binaries, presumably LMXBs, associated with the target ETGs, and (ii) the CXB. 
The radial distribution of the field-LMXBs is expected to closely follow the starlight distribution of the host galaxies (e.g., \citealp{2010ApJ...721.1368L,2013A&A...556A...9Z,min14}). 
For the latter, we adopt the S\'{e}rsic profile,
\begin{equation}
I(r) = I_{\rm e} \exp\{-b_{\rm n}[(\frac{r}{r_{\rm e}})^{1/n}-1]\},
\end{equation}
where $r_{\rm e}$ is the effective radius, $I_{\rm e}$ is the surface brightness at $r_{\rm e}$ (to be evaluated through the $g$-band absoluate magnitude $M_g$), and $b_{\rm n} \approx 2n-0.324$ for $0.5 \leq n \leq 10$ (Ciotti 1991). 
The three free parameters, $r_{\rm e}$, $n$ and $M_g$, have been derived for each of the AMUSE-Virgo galaxies in the HST/ACS Virgo Cluster Survey (ACSVCS; \citealp{2006ApJS..164..334F}), based on $g$-band photometry. 
The cumulative S\'{e}rsic profile is normalized by a factor of 5.6 sources per $10^{11} {\rm~M}_\odot$, which is derived by adopting the field-LMXB luminosity function of Zhang et al.~(2011) and taking into account our position-dependent detection limits ($\gtrsim$$2\times10^{38}{\rm~erg~s^{-1}}$).
The stellar mass of each galaxy, ranging from $5.0\times10^8{\rm~M_\odot}$ to $7.9\times10^{10}{\rm~M_\odot}$, is adopted from \cite{2008ApJ...680..154G}. 
This empirical field-LMXB contribution is plotted as a dotted curve in Figure~\ref{fig:radial}. 
We also calculate the radial dependence of the CXB contribution (dashed curve in Figure~\ref{fig:radial}), 
again correcting for the detection incompleteness and Eddington bias.

From Figure~\ref{fig:radial}, it can be seen that sources detected at $R\lesssim$0\farcm5 can be reasonably well accounted for by field-LMXBs, with a minor addition from the CXB. 
The excess at $R \lesssim 0\farcm2$ (i.e., the innermost bin) over the prediction can be understood as GC-LMXBs (see Section \ref{subsec:optical}), which have not been taken into account.
On the other hand, increasing the normalization of the field-LMXBs by a factor of two, for instance, would already overpredict the observed surface density profile between $R \approx 0\farcm3-0\farcm5$.  
Hence we conclude that the adopted field-LMXB component is reasonable.

The field-LMXB component drops rapidly beyond $R \approx 0\farcm5$.
We note that the median effective radius ($\bar{R}_e$) of the 80 AMUSE-Virgo galaxies is $0\farcm2$ (the position of 3\,$\bar{R}_e$ is marked by a vertical line in Figure~\ref{fig:radial}).
Between $0\farcm5 \lesssim R \lesssim 4^\prime$, an excess over the combined LMXB+CXB contribution is clearly present, having a relatively flat radial distribution. 
At further larger radii ($R \gtrsim 4^\prime$), the excess becomes insignificant and the predicted CXB contribution matches the observed surface density profile quite well, which is consistent with the agreement found in the flux density distribution (Figure~\ref{fig:logN}).

To be more quantitative, we estimate the significance of the excess, defined as $(N_{\rm obs}-N_{\rm LMXB}-N_{\rm CXB})/(\sqrt{N_{\rm CXB}^2 \sigma_{c}^2 +N_{\rm obs}^2 \sigma_{P}^2})$. 
Here, $\sigma_P = 1/\sqrt{N_{\rm obs}}$ is the Poisson variance for the observed number of sources ($N_{\rm obs}$). 
The cosmic variance ($\sigma_c$) of the CXB can be approximated as (\citealp{1969PASJ...21..221T}, \citealp{1992ApJ...396..430L}),
\begin{equation}
\sigma_{c}^{2}=\frac{1}{\Omega^2}\int w(\theta)d\Omega_{\rm 1}d\Omega_{\rm 2}  \approx (x^{2}+y^{2})^{-(s-1)/2}\theta_{0}^{s-1}\Theta^{1-s},
\end{equation}
where $(x^{2}+y^{2})^{1/2}\Theta$ is the angular separation between two fields, $\Theta = \Omega^{1/2}$ is the characteristic length of the field-of-view, $w(\theta)=(\theta/\theta_{0})^{s-1}$ is a power-law angular correlation function, 
and $\theta_{0}$ is the correlation length. 
We adopt the canonical value of $s=1.8$, $\Theta=\sqrt{2} \times 8\farcm4 \approx 0.198$ deg for the ACIS fields, $(x^2+y^2)^{1/2} \approx 21.1$ for the mean angular separation of 4.18 deg among the target galaxies, and $\theta_{0} \approx 0.00214 {\rm~deg}$, measured from a serendipitous {\it XMM-Newton} survey with a limiting flux of $\sim 10^{-15}{\rm~erg~s^{-1}~cm^{-2}}$ (\citealp{2009A&A...500..749E}). 
The resultant cosmic variance is $\sigma_{c} \approx 0.048$. 
Hence, for $N_{\rm obs}=601$ and the predicted values of $N_{\rm LMXB}= 33.4$ and $N_{\rm CXB}=451.6$ between $0\farcm5 < R < 4^\prime$, there are 116.0 excess sources with a significance of $\sim 3.5\,\sigma$.
In Figure~\ref{fig:s2n}, we further show the cumulative significance of excess sources as a function of projected radius, which reaches a maximum of $4.0\,\sigma$ at $R \approx 17$ kpc (3\farcm5) and gradually drops to $\sim 2.6\,\sigma$ at $R \approx$ 38 kpc ($8^\prime$), with a cumulative excess of 128.8 sources.

We have not included in the above any systematic uncertainty in $N_{\rm LMXB}$, but the contribution of field-LMXBs is negligible at large radii where the excess is concerned. 
We have also neglected the potential contribution from foreground sources. A conservative estimate, based on the detected stellar X-ray sources (including Sun-like stars, coronally active binaries and cataclysmic variables) in the Solar neighborhood \citep{saz06}, indicates that only $\lesssim$0.3 foreground sources per ACIS field (equivalent to a surface density of $\lesssim 2\times10^{-3}{\rm~arcmin^{-2}}$) are present.

As stated in Section~\ref{subsec:detection}, some of the AMUSE-Virgo fields cover more than one known member galaxies of Virgo.
This may introduce an ambiguity in the association of a given source, leading to a mistaken projected radius. 
To test this potential effect, we include the additional 23 EVCC galaxies (Table~\ref{tab:log}) to reconstruct the $F$-band source surface density profile, in which the source radius is recalculated if its projected location is closer to the center of the second galaxy than to the center of the original target galaxy.  
In this way, we find an excess of 124.7 sources with a similar significance of $\sim 3.6\,\sigma$ between $0\farcm5 < R < 4^\prime$, indicating that our result is not affected by the presence of the second galaxies.

Now we turn to investigate possible dependence of the excess sources on globel galaxy properties. 
We first divide the 80 AMUSE-Virgo galaxies into two subsets, according to their $r$-band magnitude, which is a good proxy of the stellar mass. 
The {\it brighter} subset contains 40 galaxies brighter than 13.4 mag, while the {\it fainter} subset contains the remaining 40 galaxies between 13.4-16 mag. 
In the brighter subset, we find $N_{\rm obs}=297$, $N_{\rm LMXB}= 31.8$ and $N_{\rm CXB}=213.8$ between $0\farcm5 < R < 4^\prime$, hence giving 51.4 excess sources with a significance of $\sim 2.6\,\sigma$.
In the fainter subset, $N_{\rm obs}=307$, $N_{\rm LMXB}= 2.3$ and $N_{\rm CXB}=228.7$ between $0\farcm5 < R < 4^\prime$, giving 76.0 excess sources with a significance of $\sim 3.7\,\sigma$.
This suggests, at least, no positive correlation exists between the excess and the stellar mass of the presumed central galaxies.  

We also divide the 80 fields into two subsets according to their projected distances to the cluster center, defined as the center of M\,87.
Specifically, the {\it inner} subset contains 40 galaxies located within the central $3^{\circ}$, while the {\it outer} subset contains 40 galaxies located beyond $3^{\circ}$.
In the inner subset, we find $N_{\rm obs}=263$, $N_{\rm LMXB}= 14.5$ and $N_{\rm CXB}=218.0$ between $0\farcm5 < R < 4^\prime$, hence there are 30.5 excess sources with a significance of $\sim 1.5\,\sigma$.
The outer subset has $N_{\rm obs}=338$, $N_{\rm LMXB}= 18.4$ and $N_{\rm CXB}=234.6$ between $0\farcm5 < R < 4^\prime$, giving 85.0 excess sources with a significance of $\sim 4.1\,\sigma$.
This suggests, somewhat unexpectedly, a more significant excess toward the outer cluster region. 
We caution, however, only 2 of the 80 AMUSE-Virgo fields are located within the central degree region (Figure 1). 

The above quantifications of the excess are summarized in Table 2.

\subsection{Spectral property} \label{subsec:spectral}
To further constrain the nature of the excess sources, in particular, possible contamination from ICM clumps, we examine the source spectral property. 
We focus on sources without an optical counterpart (Section~\ref{subsec:optical}), since these sources are more likely to be in Virgo rather than background AGNs. 
We also distinguish sources located in two radial ranges, $0\farcm5 < R < 4^\prime$ and $4^\prime < R < 8^\prime$, each containing $\sim$100 sources.  
Using the CIAO tool {\it specextract}, we extract spectra from within the 90\% EER of each source of interest and build the Response Matrix Files (RMFs) and Ancillary Response Files (ARFs); the corresponding background spectrum is extracted typically between 2-5 times the 90\% EER. 
We then coadd the individual source spectra to generate a cumulative spectrum, with weighted RMFs and ARFs, for each of the two radial ranges.
We note that the cumulative spectrum is not dominated by any few sources.

As shown in Figure~\ref{fig:spectra}, the cumulative spectra appear 
featureless and can be well characterized by an absorbed power-law model.
We find that the absorption column density, when left as a free parameter, is consistent with the Galactic foreground absorption, $N_{\rm H} \approx 2.5\times10^{20}{\rm~cm^{-2}}$. Fixing the absorption at this value, we obtain the best-fit photon-index, $1.56\pm0.07$ ($1.70\pm0.06$) for sources located at $0\farcm5 < R < 4^\prime$ ($4^\prime < R < 8^\prime$), which is consistent with accretion-powered soruces, i.e., LMXBs and/or AGNs. 
If the spectra were instead fitted with an absorbed thermal plasma model (APEC in XSPEC), we obtain a best-fit plasma temperature of $\sim$5-8 keV. This is several times higher than the characteristic temperature of the Virgo ICM ($\sim$2 keV, Urban et al.~2011). 
It is highly unlikely that ICM clumps of such a high temperature can be so commonly formed and pressure-confined inside Virgo.  
Hence we conclude that the detected X-ray sources have a discrete nature, and that contamination from ICM clumps, if any, is negligible.

\subsection{Optical Counterparts} \label{subsec:optical}
We search for optical counterparts of all detected X-ray sources, using the source catalog from the ACSVCS \citep{jor09} and the unpublished catalog from Next Generation Virgo Cluster Survey (NGVS) on the Canada-France-Hawaii Telescope \citep{fer12}. The ACSVCS has high-resolution and high-sensitivity for resolving compact sources such as GCs, but its field-of-view is limited.
The NGVS has the complementary advantage of wide area essentially covering all the AMUSE-Virgo fields, and is also sensitive to detecting GCs and background galaxies. 
We adopt a matching radius of twice the X-ray source position error (typical value $\sim 0\farcs9$) when cross-correlating the X-ray and optical catalogs. If a counterpart was found in both the ACSVCS and NGVS catalogs, only one pair is counted. A total of 887 pairs are thus identified, with less than 0.1\% random matches. 
We have estimated the latter value by artificially shifting the X-ray positions by $\pm10^{\prime\prime}$ in RA and DEC and averaging the resultant number of coincident matches. 

In Figure \ref{fig:radial}b, we plot the average surface density profile in red histogram for those X-ray sources with an optical counterpart. We find that 787/1047 = 75.2\% of the $F$-band sources have an optical counterpart (317/413 = 76.8\%, if only sources located beyond $4^\prime$ were considered), which follow closely with the predicted CXB distribution at all radii. This is consistent with the vast majority of the X-ray sources being background galaxies/AGN, which can have a $\sim$80\% optical identification rate in deep surveys (e.g., \citealp{luo10}).

We point out that 8 optical counterparts are found within the innermost bin in Figure \ref{fig:radial}, among which 7 are classified as GCs in the ACSVCS catalog, nicely explaining the apparent excess there.
We further identify 20 ACSVCS GCs each positionally coincident with an X-ray source located beyond $R = 0\farcm5$.
An addition of $\sim$15 GCs, preliminarily classified in the NGVS images, are found to have an X-ray counterpart.
The potential contribution of GC-LMXBs to the excess will be further discussed in Section \ref{sec:discussion}.
We defer a detailed analysis of all optical counterparts to a future work. 

\subsection{X-ray sources in the control fields} \label{subsec:field}
Figure~\ref{fig:field} shows the average surface density profile of the 868 $F$-band sources detected within a projected radius of $8^\prime$ of the 57 AMUSE-Field galaxies. 
We also plot the predicted field-LMXB and CXB contributions in these galaxies. 
For the LMXB component, we again adopt the cumulative S\'{e}rsic profile, making use of the empirical $r_{e}-M_{\rm B}$ and $n-M_{\rm B}$ relations obtained from Virgo ETGs (Ferrarese et al.~2006; Equations 26 and 27 therein) and B-band magnitudes of the AMUSE-Field galaxies from Miller et al.~(2012b).
The median effective radius of these field galaxies is $\bar{R}_{\rm e} \approx 0\farcm13$. 
Within $\sim3\bar{R}_{\rm e}$ (marked by a vertical line in Figure~\ref{fig:field}), the field-LMXBs, plus a minor contribution from the CXB, provide a good match to the observed profile.   
Little excess can be seen at larger radii, except for a small ``bump" at $R \approx 1^\prime - 2^\prime$. 
We note that, while dubbed ``field galaxies", some of the AMUSE-Field galaxies are actually located in group environments (Miller et al.~2012b), and hence the observed profile might be contaminated by sources physically associated with neighboring galaxies of the same group. 
This seems particularly relevant to the ``bump".
In any case, the presence of some halo sources in field galaxies is expected, as demonstrated by Li et al.~(2010) and Zhang et al.~(2013).

\begin{figure*}\centering
\includegraphics[width=\textwidth,angle=90]{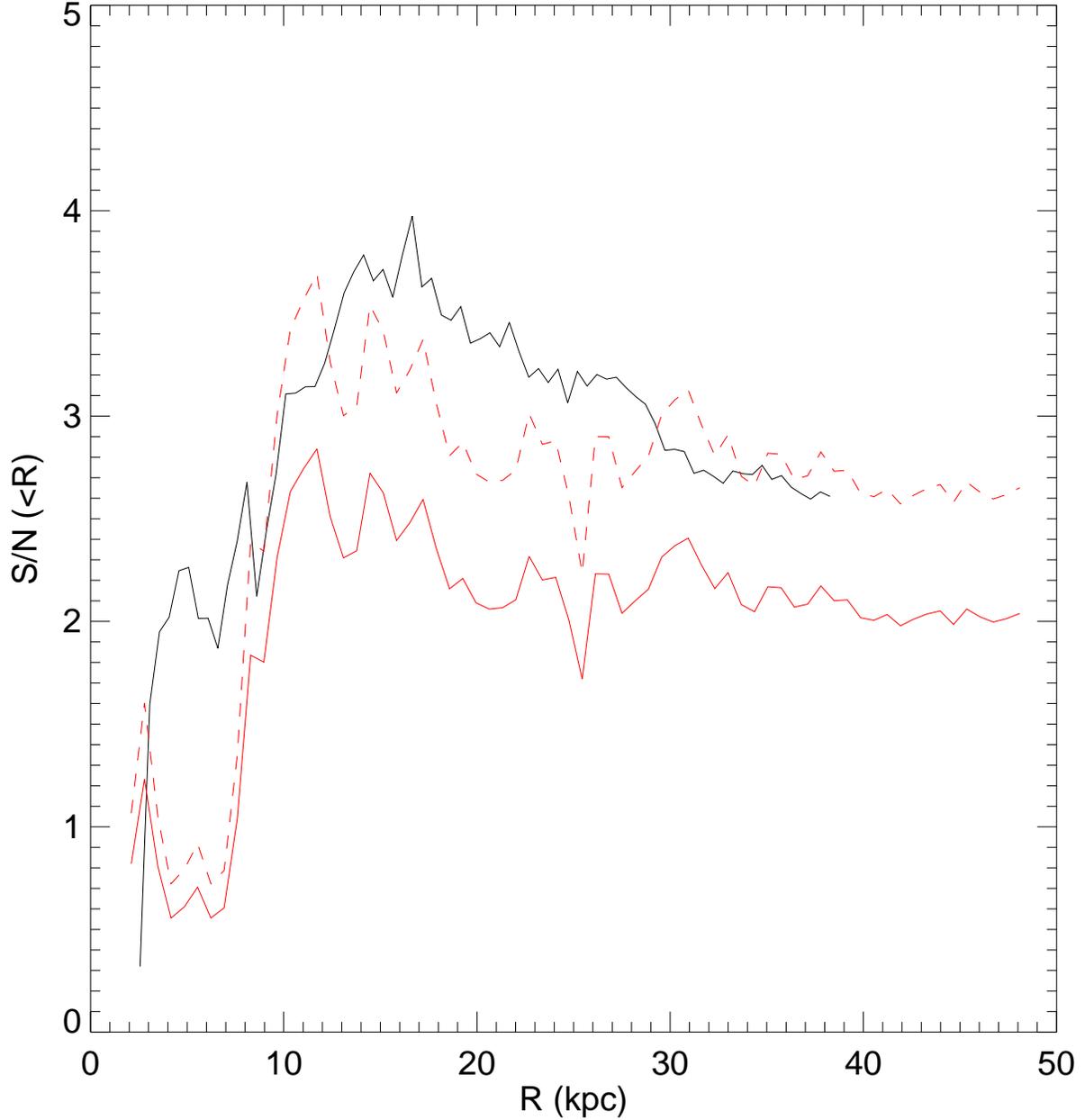}
\caption{The significance of $F$-band excess sources over the predicted sum of field-LMXBs and CXB, as a function of projected radius from the target galaxies.
The projected radius of the AMUSE-field galaxies (red curves) has been converted into units of kpc by adopting their median distance of 22.0 Mpc, to facilitate a more direct comparison with the AMUSE-Virgo galaxies (black curve).
The solid curves are derived following $(N_{\rm obs}-N_{\rm LMXB}-N_{\rm CXB})/(\sqrt{N_{\rm CXB}^2 \sigma_{c}^2 +N_{\rm obs}^2 \sigma_{P}^2})$. 
The dashed red curve is derived by further multiplying a factor of $\sim$1.3, which is the ratio of the predicted CXB surface densities between AMUSE-Virgo and AMUSE-Field, to compensate for the higher CXB level in the latter due to its on-average longer exposures. 
See text for details.
}
\label{fig:s2n}
\end{figure*}

\begin{figure*}\centering
\includegraphics[scale=0.65, angle=270]{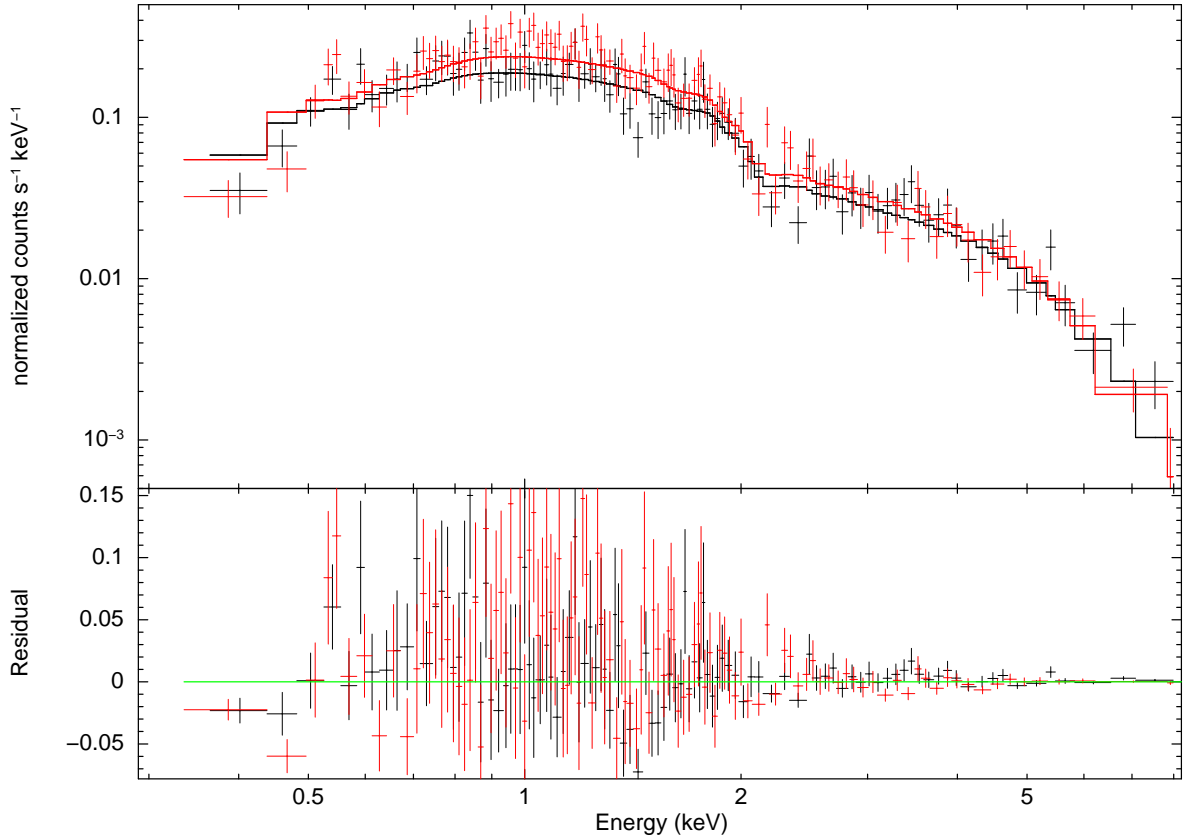}
\caption{{The cumulative spectra of sources without an optical counterpart, black for $0\farcm5 < R < 4^\prime$ and red for $4^\prime < R < 8^\prime$. The fitted model is an absorbed power-law with a photon-index of $1.56\pm0.07$ and $1.70\pm0.06$, respectively.
The lower panel shows the residual (data $-$ model).}
}
\label{fig:spectra}
\end{figure*}

\begin{figure*}\centering
\includegraphics[width=\textwidth,angle=90]{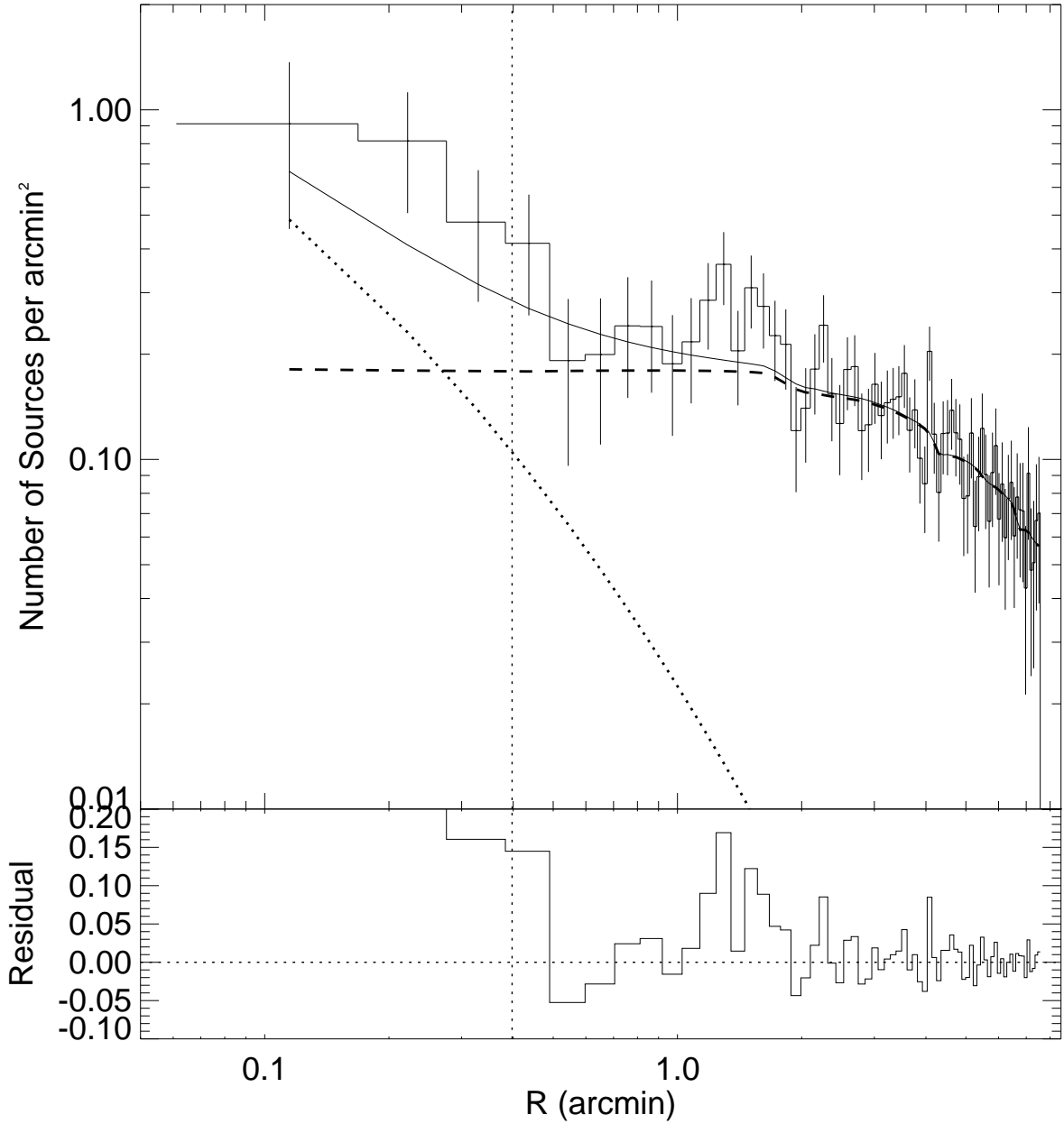}
\caption{Radial surface density distribution of $F$-band sources, averaged over the AMUSE-field galaxies. 
The dotted curve represents the expected contribution of field-LMXBs associated with the target galaxies.
The dashed curve represents the cosmic X-ray background, corrected for detection incompleteness and Eddington bias.
The vertical dashed line marks the position of $R=3R_{\rm e}$, where $R_{\rm e} = 0\farcm13$ is the median effective radius of all target galaxies.
The lower panel shows the residual after subtracting the combined contribution (solid curve) of field-LMXBs and CXB.
}
\label{fig:field}
\end{figure*}

The median distance of the AMUSE-Field galaxies is 22.0 Mpc.
Thus a projected radial range of $0\farcm3-2\farcm5$ appears a fair zone for comparison with the range of $0\farcm5-4\farcm0$ ($\sim3-20$ kpc) in the AMUSE-Virgo galaxies, where the most significant excess is found (Section \ref{subsec:spatial}).
 Within this range of the field galaxies, a total of 242 $F$-band sources are detected, while the predicted number of field-LMXBs and CXB sources is 15.8 and 188.5, respectively. 
This only gives an excess of 37.7 sources, with a significance of $\sim 2.4\,\sigma$ (versus 3.5\,$\sigma$ in the AMUSE-Virgo galaxies; Section \ref{subsec:spatial}).
The cumulative significance of the excess sources in the AMUSE-Field galaxies as a function of projected radius is shown as a red solid curve in Figure \ref{fig:s2n}. 
It is noteworthy that the CXB level in AMUSE-Field is by a factor of $\sim$1.3 higher than in AMUSE-Virgo, due to the on-average longer exposures in the former in order to achieve the same limiting luminosity (Section 2.1). This results in a higher ``noise" for the excess sources. Hence we recalculate the cumulative significance of excess by dividing the ``noise" by a factor of 1.3 to compensate for this effect (dashed red curve in Figure \ref{fig:s2n}). 
Still, it is clear that over the range of $3-30$ kpc, excess in the field galaxies, if any, is substantially lower than found in the AMUSE-Virgo galaxies. 



\section{Discussion} \label{sec:discussion}
In the above we have surveyed the vicinity of 80 ETGs in Virgo using {\it Chandra} observations, and gathered strong evidence for the presence of $\sim$120 excess sources with respect to the predicted sum of field-LMXBs and CXB, over a projected radial range of $3-30$ kpc.   
From now on, we shall refer to this excess, collectively, as intra-cluster X-ray sources, to underscore the fact that they are located beyond the main stellar content of the target galaxies.
We caution that some, if not most, of these excess sources could still be physically associated with, rather than gravitationally unbound from, the target galaxies. 
Firmly establishing the latter case requires kinematic information, which is not yet available.
Nevertheless, just as the diffuse intra-cluster light can be partially bound to individual member galaxies and there is no clear dividing line (Mihos 2015), it is not unreasonable to use here the terminology of intra-cluster X-ray sources. 
Moreover, no significant excess sources are found around the AMUSE-Field galaxies, strongly suggesting that at least a good fraction of the excess sources in Virgo are uniquely related to the cluster environment.
Below, we consider possible origins for these sources, including     
(i) GC-LMXBs, (ii) supernova-kicked LMXBs, (iii) LMXBs associated with the ICL, (iv) stripped galactic nuclei, and (v) gravitationally recoiled MBHs.
The first two scenarios, discussed in Li et al.~(2010) and Zhang et al.~(2013), are more appropriate for halo X-ray sources bound to the host galaxy, 
whereas the latter three cases are most relevant to the cluster environment and more likely to represent truly intra-cluster X-ray populations.

\subsection{GC-LMXBs and supernova-kicked LMXBs}
Statistically, a few percent of the extragalactic GCs are found to host an X-ray counterpart with $L_{\rm X} \gtrsim 10^{37}{\rm~erg~s^{-1}}$ (Fabbiano 2006; Kim et al.~2009,2013; Hou \& Li 2016). 
It is well known that the radial distribution of GCs, in particular blue GCs, is substantially broader than that of the stellar spheroid. 
Hence no doubt that some of the intra-cluster X-ray sources are GC-LMXBs.
Indeed, 35 GC candidates are found to be positionally coincident with the X-ray sources beyond $R = 0\farcm5$ (Section \ref{subsec:optical}), thus making up of $\sim$30\% of the excess.
However, GC-LMXBs are unlikely to account for the bulk of the intra-cluster X-ray sources.  
Empirically, the vast majority of GC-LMXBs are found in massive GCs that are above the turnover magnitude (e.g., Li et al.~2010). The ACSVCS and NGVS images have sensitivities reaching well above the GC turnover magnitude \citep{jor05}, hence it is unlikely that a large fraction of GC hosts are missed.  

Core-collapsed supernova explosions, if occurring in binary systems, can give rise to a substantial kick velocity in the surviving binary which now contains a compact object, favorably a neutron star. 
If this kick velocity is comparable to or exceeds the escape velocity (on the order of 100 km~s$^{-1}$) of the host galaxy, a LMXB system might be later found in the halo (Brandt \& Podsiadlowski 1995; Zuo et al.~2008). 
Zhang et al.~(2013; see also Li et al.~2010) proposed that a good fraction of the halo X-ray sources found in their sample ETGs can be such supernova-kicked LMXBs, for the number of such sources scale with stellar mass of the host galaxy.
Supernova-kicked LMXBs should also be present among our intra-cluster X-ray sources, and their optical counterpart is probably too faint to be resolved.  
However, we suggest that they cannot be a dominant component, either, for two reasons.
First, we find no positive correlation between the significance of the excess and the stellar mass of the putative host galaxies (Section \ref{subsec:spatial}). 
Second, and more importantly, we recall that the limiting luminosity in all AMUSE-Virgo fields ($\gtrsim$$3\times10^{38}{\rm~erg~s^{-1}}$ between $0\farcm5-4^\prime$; Figure \ref{fig:logN}) exceeds the Eddington limit of neutron star-LMXBs. Therefore, only a small fraction of the detected intra-cluster sources should be neutron star-LMXBs\footnote{This argument is also relevant to the X-ray sources associated with GCs, although the total X-ray luminosity of a GC may be due to a superposition of multiple neutron star binaries or from a single, rare black hole binary \citep{mac07}.}. 
In fact, Zhang et al.~(2013) found no significant halo excess at $L_{\rm X} \gtrsim 5\times10^{38}{\rm~erg~s^{-1}}$, whereas our Virgo sources above the same luminosity threshold still show a 2.1\,$\sigma$ excess (Table 2), although this difference may be due to the much smaller ETG sample of Zhang et al. compared to ours (20 versus 80).  
Nevertheless, we cannot rule out the possibility that supernova-kicked black hole binaries exist among the intra-cluster X-ray sources, although the probability for the supernova explosion to boost these heavier systems should be much lower. 
Alternatively, some supernova-kicked neutron stars might have later collapsed into black holes via accretion from the companion star \citep{bra95b}.

\subsection{LMXBs associated with the ICL}
The presence of diffuse ICL in Virgo has long been established \citep{mih05}, which is believed to be the result of frequent tidal encounters and mergers among its thousands of member galaxies. 
The red color of the ICL implies for predominantly old stellar populations \citep{mih17}, which naturally form LMXBs (ICL-LMXBs).
That a more significant excess is found around fainter (smaller) galaxies (Table 2) is qualitatively consistent with such a scenario.
We can have a rough estimate of the number of ICL-LMXBs falling within the AMUSE-Virgo fields, as $N$(ICL-LMXB) $= \epsilon f_{\rm FoV} f_{\rm ICL} f_{\rm stars} M_{200}$. 
Here $M_{200} \approx 2\times10^{14}{\rm~M_\odot}$ is the virial mass of Virgo, estimated based on the $M_{200}-T$ relation of \cite{ket15} and a temperature of $\sim$2 keV for the hot intra-cluster medium (ICM; \citealp{urb11});
$f_{\rm stars} \approx 0.05$ is the fractional mass in stars empirically measured for clusters as massive as Virgo \citep{gon07}, and $f_{\rm ICL} \approx 10\%$ is the fraction of stars in the ICL \citep{mih17}.
The fractional coverage of the ICL by the 80 AMUSE-Virgo fields, $f_{\rm FoV}$, is $\sim$10\%, given the total number (973) of EVCC galaxies brighter than 16 mag in the $r$-band (Section \ref{subsec:detection}).  
Lastly, the abundance of LMXBs above a mean detection limit of $3\times10^{38}{\rm~erg~s^{-1}}$, $\epsilon \approx 10$ per $10^{11}{\rm~M_\odot}$, is derived assuming the same linear relation with the underlying stellar mass as found in ETGs (Zhang et al.~2011).
The above estimate thus predicts $\sim$10 ICL-LMXBs. 
 Admittedly, the above parameters have their own uncertainties, which can be up to 100\%, e.g., in $M_{200}$, but at face value the ICL-LMXBs are unlikely to account for the majority of the intra-cluster X-ray sources.
For the ICL-LMXBs to have a larger contribution, one possibility is to have a flatter X-ray luminosity function (thus a higher value of $\epsilon$), as predicted in the presence of relatively young ($\sim$ 1 Gyr) stellar populations (Wu 2001; Fragos et al.~2008).
However, the star formation history in the ICL is currently poorly understood.

\subsection{Relic galactic nuclei and recoiled MBHs}
The advent of high-resolution optical imaging surveys in the past decades has accmulated strong evidence for the 
formation of tidally stripped nucleated dwarf galaxies in cluster environments (e.g., Ferguson \& Binggeli 1994; Ferrarese et al.~2016). 
Closely related are the so-called ultra-compact dwarfs (UCDs; \citealp{phi01,bru12}), which are compact stellar systems with luminosities and sizes (hence densities) intermediate between classical GCs and dwarf elliptical galaxies. 
At least a subset of the UCDs share a common origin with the nucleated dwarf galaxies (e.g., \citealp{liu15,fer16}).
Hou \& Li (2016; see also \citealp{pan16}) found that about 3\% of the known UCDs have an X-ray counterpart (with $L_{\rm X} \gtrsim 10^{37}{\rm~erg~s^{-1}}$), which they suggested to be predominantly LMXBs having formed in stellar dynamical interactions. 
In particular, Hou \& Li (2016) detected an X-ray counterpart in M60-UCD1 (already found by Strader et al. 2013) and M59-UCD3, two of the most massive UCDs currently known, and another 5 less massive UCDs in NGC\,4365. These UCDs reside in Virgo, although they are not covered by the AMUSE-Virgo fields. This clearly suggests that UCDs can contribute to the intra-cluster X-ray sources.
An extensive identification of UCDs using the NGVS data is underway (C. Liu et al. in preparation).

Some UCDs, especially the massive ones, may harbor a central MBH. This is the case in M60-UCD1 \citep{set14}, in which a central black hole with $M_{\rm BH} \approx 2\times10^7 {\rm~M_\odot}$ was found embedded in $10^{8}{\rm~M_\odot}$ of stars. It is plausible that the X-ray source in M60-UCD1, and by analogy in other X-ray-detected UCDs, is powered by the central MBH.  

Another form of MBHs wandering in the intra-cluster space could be the long-sought gravitationally recoiled MBHs after the merger of two progenitor MBHs \citep{sch13}. 
A naked recoiled MBH having escaped the parent galaxy might be powered by Bondi accretion from the ICM, at a rate of (e.g., Li et al.~2011), 
\begin{equation}
\dot{M}_{\rm Bondi} \approx 4\pi \lambda \mu m_{H}n_{\rm ICM}(GM_{\rm BH})^{2}c_{s}^{-3} \approx 2.2\times 10^{-6}(\frac{n_{\rm ICM}}{10^{-3}{\rm~cm^{-3}}})(\frac{kT}{\rm 2~keV})^{-1.5}(\frac{M_{\rm BH}}{10^9{\rm~M_\odot}})^{2}{\rm~M_\odot~yr^{-1}},
\end{equation}
where the characteristic density ($n_{\rm ICM}$) and temperature ($T$) of the ICM are from \cite{urb11}. 
Assuming a radiation efficiency $\eta \approx 10\%$ and that 10\% of the bolometric luminosity goes to the X-ray band, to have $L_{\rm X} \approx 0.01 (\eta/0.1) \dot{M}_{\rm Bondi} c^{2} \gtrsim 3\times10^{38}{\rm~erg~s^{-1}}$, 
a black hole mass of $M_{\rm BH} \gtrsim 5\times10^{8}{\rm~M_\odot}$ is needed. 
Since the total stellar mass in Virgo is about $\sim10^{13}{\rm~M_\odot}$ (Section 4.2), it seems unlikely to have more than $\sim$100 such SMBHs wandering in the cluster ($\sim$10 captured by AMUSE-Virgo). 
Alternatively, if there exist cooler and denser gaseous filaments in the intra-cluster space, e.g., due to ram pressure stripping of cold gas from late-type galaxies, a MBH with $10^{5-6}{\rm~M_\odot}$ occasionally runs into such a filament could be easily lit up to the observed X-ray luminosities for thousands to a million years.
Only under such fortunate circumstance can we expect naked MBHs to behave as intra-cluster X-ray sources.

In reality, the recoiled MBH may be surrounded by a relic cluster of stars which it pulls out from the nucleus of the parent galaxy. 
To our knowledge, there has been no theoretical work dedicated to predict the X-ray luminosities of such recoiled MBHs. 
Using the simple modeling of Volonteri et al.~(2011), we estimate that a MBH of $\gtrsim10^{6}{\rm~M_\odot}$, under Bondi accretion from stellar winds in the relic star cluster, can reach $L_{\rm X} \sim 10^{38}{\rm~erg~s^{-1}}$. 
Interestingly, such systems can have a similar appearance as GCs or UCDs in terms of size and luminosity, but with much larger velocity dispersions due to the presence of the central MBH (Merritt et al. 2009). 
This invokes optical spectroscopic follow-up observations toward the intra-cluster X-ray sources found in this work. 

We suggest that relic galactic nuclei and recoiled MBHs can have a non-neglibile, albeit unlikely dominant, contribution to the intra-cluster X-ray sources.

\section{Conclusions} \label{sec:sum}
Using archival {\it Chandra} observations, we have surveyed the vicinity of 80 ETGs in the Virgo cluster and detected $\sim$1200 point sources outside the necleus of these galaxies. After examining the cumulative spatial and flux distributions, we find that $\sim$10\% of these sources can be statistically classified as intra-cluster X-ray sources, i.e., not associated with the bulk stellar content of the putative host galaxies, nor with the CXB. 
We suggest that these intra-cluster X-ray sources have a mixed origin, including: 

\begin{itemize} 
\item GC-LMXBs, which might contribute up to $\sim$30\% of the intra-cluster sources; 
\item Supernova-kicked LMXBs, which, however, are not expected to have a significant contribution; 
\item ICL-LMXBs, which might contribute to 10\% or more;
\item Relic galactic nuclei and gravitationally recoiled MBHs, the contribution of which is not readily strongly constrained, but confirming the existence of such objects as intra-cluster X-ray sources would have important implications on galaxy evolution inside the cluster environment. 
\end{itemize}

In a subsequent work we will present the optical properties of the X-ray sources, based on imaging and spectroscopic follow-up observations. 
Future X-ray observations covering a large area of Virgo and reaching luminosities well below $10^{38}{\rm~erg~s^{-1}}$ would be useful to shed light on the nature of the intra-cluster populations.

\begin{acknowledgements}
This work is supported by the National Science Foundation of China under grant 11133001.
We acknowledge the PIs of the AMUSE-Virgo and AMUSE-Field programs, Dr. T. Treu and Dr. E. Gallo, for acquiring the data that make this work possible. 
We thank Zhongli Zhang for helpful discussions. 
M.H. is grateful to the hospitality of KIAA/PKU during her visit.
Z.L. acknowledges support from the Recruitment Program of Global Youth Experts.

\end{acknowledgements}

\begin{deluxetable}{cccccccccccc}
\tabletypesize{\scriptsize}
\tablecaption{Basic information of the Virgo galaxies}
\tablewidth{0pt}
\tablehead{
\colhead{Galaxy name} &
\colhead{Other name} &
\colhead{ObsID} &
\colhead{RA} &
\colhead{DEC} &
\colhead{Exp.} &
\colhead{$L_{\rm X, lim}$} &
\colhead{${N_{\rm obs}}$} &
\colhead{${N_{\rm CXB}}$} &
\colhead{Second galaxy}\\
\colhead{(1)} &
\colhead{(2)} &
\colhead{(3)} &
\colhead{(4)} &
\colhead{(5)} &
\colhead{(6)} &
\colhead{(7)} &
\colhead{(8)} &
\colhead{(9)} &
\colhead{(10)} 
}
\startdata
VCC\,9&		IC\,3019& 8072& 182.342748& 13.992428& 5.3& 1.9& 9& 10.9\\ 
VCC\,21&	IC\,3025& 8089& 182.596211& 10.188544& 5.5& 1.8& 18& 11.0\\
VCC\,33&	IC\,3032& 8086& 182.782342& 14.274814& 5.2& 1.9& 8& 10.9\\
VCC\,140&	IC\,3065& 8076& 183.802344& 14.432881& 5.1& 2.0& 24& 10.2\\ 
VCC\,200& --&            8087& 184.140440& 13.031573& 5.1& 2.0& 14& 10.4\\
VCC\,230& IC\,3101& 8100& 184.331885& 11.943467& 5.1& 2.0& 13& 10.3\\ 
VCC\,355& NGC\,4262& 8049& 184.877395& 14.877655& 4.7& 2.6& 15& 7.0\\
VCC\,369& NGC\,4267& 8039& 184.938482& 12.798283& 5.1& 2.0& 14& 10.2\\ 
VCC\,437& UGC\,7399A& 8085& 185.203411& 17.487065& 5.1& 2.0& 11& 10.5\\ 
VCC\,538& NGC\,4309A& 8105& 185.561425& 7.167147& 5.3& 1.9& 16& 10.7& VCC534\\
VCC\,543& UGC\,7436& 8080& 185.581370& 14.760780& 5.3& 1.9& 16& 10.7\\
VCC\,571& --& 8088& 185.671491& 7.950348& 5.1& 2.0& 11& 10.4&  VCC584\\
VCC\,575& NGC\,4318& 8073& 185.680393& 8.198277& 5.1& 2.0& 20& 10.4\\
VCC\,654& NGC\,4340& 8045& 185.897026& 16.722359& 5.1& 2.0& 14& 10.4\\
VCC\,698& NGC\,4352& 8068& 186.020937& 11.218066& 5.1& 2.0& 11& 10.3\\
VCC\,751& IC\,3292& 8103& 186.201501& 18.195123& 5.1& 2.0& 16& 10.5\\
VCC\,759& NGC\,4371& 8040& 186.230960& 11.704210& 4.9& 2.1& 17& 9.8& VCC786\\
VCC\,778& NGC\,4377& 8055& 186.301400& 14.762160& 5.1& 2.0& 18& 10.1\\
VCC\,784& NGC\,4379& 8053& 186.311417& 15.607472& 5.1& 2.0& 21& 10.2\\
VCC\,828& NGC\,4387& 8056& 186.423668& 12.810515& 5.1& 2.0& 13& 8.6\\
VCC\,856& IC\,3328& 8128& 186.491360& 10.053767& 5.1& 2.0& 8& 10.3\\
VCC\,944& NGC\,4417& 8125& 186.710880& 9.584260& 5.1& 2.0& 15& 10.1\\
VCC\,1025& NGC\,4434& 8060& 186.902847& 8.154344& 5.1& 2.0& 13& 10.2\\
VCC\,1030& NGC\,4435& 8042& 186.918698& 13.078945& 4.9& 2.2& 22& 7.8& VCC1043\\
VCC\,1049& UGC\,7580& 8075& 186.978467& 8.090403& 5.5& 1.8& 12& 11.0& VCC1019\\
VCC\,1062& NGC\,4442& 8037& 187.016180& 9.803710& 5.3& 1.9& 26& 10.4& VCC1078\\
VCC\,1075& IC\,3383& 8096& 187.051343& 10.297657& 5.1& 2.0& 17& 10.0\\
VCC\,1087& IC\,3381& 8078& 187.062016& 11.789826& 5.1& 2.0& 13& 9.4\\
VCC\,1125& NGC\,4452& 8064& 187.180455& 11.755032& 5.1& 2.0& 16& 9.3& VCC1115\\
VCC\,1146& NGC\,4458& 8059& 187.239855& 13.241893& 5.1& 2.0& 11& 9.1& VCC1158\\
VCC\,1178& NGC\,4464& 8127& 187.338716& 8.156623& 5.1& 2.4& 1& 8.1\\
VCC\,1185& --& 8110& 187.347975& 12.450786& 5.1& 2.4& 10& 7.5& VCC1213\\
VCC\,1242& NGC\,4474& 8052& 187.473113& 14.068589& 5.1& 2.4& 6& 7.7\\
VCC\,1250& NGC\,4476& 8057& 187.496172& 12.348657& 5.1& 2.8& 11& 5.5\\
VCC\,1261& NGC\,4482& 8067& 187.543029& 10.779485& 5.1& 2.0& 17& 10.1\\
VCC\,1283& NGC\,4479& 8066& 187.576553& 13.577643& 5.1& 2.0& 16& 9.4& VCC1253\\
VCC\,1303& NGC\,4483& 8061& 187.669357& 9.015678& 5.1& 2.0& 17& 10.3\\
VCC\,1321& NGC\,4489& 8126& 187.717706& 16.758854& 5.1& 2.0& 19& 10.2\\
VCC\,1355& IC\,3442& 8077& 187.834137& 14.115204& 5.1& 2.0& 9& 9.9\\
VCC\,1407& IC\,3461& 8131& 188.011399& 11.890073& 5.1& 2.2& 8& 7.8& VCC1411\\
VCC\,1422& IC\,3468& 8069& 188.059220& 10.251462& 5.1& 2.0& 12& 10.0\\
VCC\,1431& IC\,3470& 8081& 188.097449& 11.262968& 5.1& 2.4& 14& 7.6& VCC1412\\
VCC\,1440& IC\,798& 8099& 188.139191& 15.415390& 5.5& 1.8& 16& 12.2\\
VCC\,1475& NGC\,4515& 8065& 188.270728& 16.265534& 5.1& 2.0& 17& 11.6\\
VCC\,1488& IC\,3487& 8090& 188.305983& 9.397355& 5.1& 2.0& 18& 11.3& VCC1509\\
VCC\,1489& IC\,3490& 8113& 188.307934& 10.928536& 5.1& 2.0& 5& 11.1\\
VCC\,1499& IC\,3492& 8093& 188.332407& 12.853435& 5.1& 2.0& 20& 10.7& VCC1491\\
VCC\,1512& --& 8112& 188.394404& 11.261955& 5.1& 2.0& 12& 10.9\\
VCC\,1528& IC\,3501& 8082& 188.465078& 13.322443& 5.1& 2.0& 10& 11.2\\
VCC\,1537& NGC\,4528& 8054& 188.525309& 11.321259& 5.1& 2.0& 16& 10.9\\
VCC\,1539& IC\,3506& 8109& 188.528081& 12.741598& 5.1& 2.0& 15& 10.9\\
VCC\,1545& IC\,3509& 8094& 188.548060& 12.048964& 5.1& 2.0& 16& 10.6\\
VCC\,1619& NGC\,4550& 8050& 188.877428& 12.220824& 5.1& 2.0& 12& 11.0& VCC1630\\
VCC\,1627& --& 8098& 188.905215& 12.382027& 5.1& 2.0& 15& 11.0\\
VCC\,1630& NGC\,4551& 8058& 188.908144& 12.263978& 5.5& 1.8& 10& 11.8& VCC1619\\
VCC\,1661& --& 8044& 189.103269& 10.384656& 4.9& 2.1& 12& 11.2\\
VCC\,1692& NGC\,4570& 8041& 189.222500& 7.246639& 5.1& 2.0& 18& 11.8\\
VCC\,1695& IC\,3586& 8083& 189.228527& 12.520074& 5.1& 2.0& 14& 11.3\\
VCC\,1720& NGC\,4578& 8048& 189.377340& 9.55507& 5.1& 2.0& 16& 11.9\\
VCC\,1743& IC\,3602& 8108& 189.528295& 10.082354& 5.1& 2.0& 8& 11.5\\
VCC\,1779& IC\,3612& 8091& 189.769597& 14.731127& 5.1& 2.0& 20& 11.7\\
VCC\,1826& IC\,3633& 8111& 190.046876& 9.896128& 5.1& 2.0& 21& 11.6& VCC1825\\
VCC\,1828& IC\,3635& 8104& 190.055731& 12.874775& 5.1& 2.0& 20& 11.5\\
VCC\,1833& --& 8084& 190.081986& 15.935282& 5.2& 1.9& 23& 12.2\\
VCC\,1857& IC\,3647& 8130& 190.221306& 10.475364& 5.1& 2.0& 12& 11.8\\
VCC\,1861& IC\,3652& 8079& 190.244001& 11.184487& 5.1& 2.0& 14& 11.5& VCC1870\\
VCC\,1871& IC\,3653& 8071& 190.315559& 11.387245& 5.1& 2.0& 12& 11.6& VCC1870\\
VCC\,1883& NGC\,4612& 8051& 190.386463& 7.314882& 5.1& 2.0& 25& 11.8\\
VCC\,1886& IC\,3663& 8106& 190.414221& 12.247391& 5.1& 2.0& 17& 11.5\\
VCC\,1895& UGC\,7854& 8092& 190.466583& 9.402889& 5.1& 2.0& 17& 11.8\\
VCC\,1910& IC\,809& 8074& 190.536104& 11.754286& 5.3& 1.9& 26& 11.8& VCC1903\\
VCC\,1913& NGC\,4623& 8062& 190.544542& 7.676944& 5.3& 1.9& 18& 12.2\\
VCC\,1938& NGC\,4638& 8046& 190.697602& 11.442507& 5.3& 1.9& 22& 11.9& VCC1945\\
VCC\,1948& IC\,3693& 8097& 190.741746& 10.681811& 5.1& 2.0& 12& 11.5\\
VCC\,1993& --& 810& 191.050114& 12.941839& 5.1& 2.0& 12& 11.6\\
VCC\,2000& NGC\,4660& 8043& 191.133266& 11.190533& 5.1& 2.0& 22& 11.4& VCC2002\\
VCC\,2019& IC\,3735& 8129& 191.335100& 13.692662& 5.1& 1.9& 12& 11.6\\
VCC\,2048& IC\,3773& 8070& 191.813746& 10.203585& 5.3& 1.9& 19& 12.2& VCC2045\\
VCC\,2050& IC\,3779& 8101& 191.835983& 12.166427& 5.1& 2.0& 12& 11.7\\
VCC\,2092& NGC\,4754& 8038& 193.072900& 11.313886& 5.0& 2.0& 17& 10.2\\
\enddata
\tablecomments{(1) Name of the target galaxies in the AMUSE-Virgo observations; (2) Other name of the target galaxies; (3) {\it Chandra} observation ID; (4)-(5): Celestial coordinates of the galactic center (J2000); (6) {\it Chandra} effective exposure, in units of ks; (7) 0.5-8 keV limiting luminosity for source detection, in units of ${10^{38}\rm~erg~s^{-1}}$; (8) Number of X-ray sources detected in each field; (9) Number of cosmic X-ray background sources predicted by the log$N$-log$S$ relation of Georgakakis et al.~(2008); (10) Additional Virgo galaxy located within the field-of-view. 
}
\label{tab:log}
\end{deluxetable}

\begin{deluxetable}{ccccccc}
\tabletypesize{\footnotesize}
\tablecaption{Excess X-ray sources in different subsets}
\tablewidth{0pt}
\tablehead{
\colhead{Subset} &
\colhead{Radial range} &
\colhead{${N_{\rm obs}}$} &
\colhead{${N_{\rm LMXB}}$} &
\colhead{${N_{\rm CXB}}$} &
\colhead{${N_{\rm excess}}$} &
\colhead{Significance} \\
\colhead{(1)} &
\colhead{(2)} &
\colhead{(3)} &
\colhead{(4)} &
\colhead{(5)} &
\colhead{(6)} &
\colhead{(7)}
}
\startdata
All & $0\farcm5 - 8^\prime$& 965& 41.1& 795.2& 128.7& $ 2.6\,\sigma$\\ 
Fiducial  & $0\farcm5 - 4^\prime$& 601& 33.4& 451.6& 116.0& $ 3.5\,\sigma$\\ 
With second EVCC & $0\farcm5 - 4^\prime$& 629 & 31.3 & 473.0 & 124.7 & $3.6\,\sigma$\\
Brighter galaxies& $0\farcm5 - 4^\prime$&  297& 31.8& 213.8& 51.4& $ 2.6\,\sigma$\\ 
Fainter galaxies& $0\farcm5 - 4^\prime$& 307& 2.3& 228.7& 76.0& $ 3.7\,\sigma$\\ 
Inner cluster & $0\farcm5 - 4^\prime$& 263& 14.5& 218.0& 30.5& $ 1.5\,\sigma$\\ 
Outer cluster & $0\farcm5 - 4^\prime$&338& 18.4& 234.6& 85.0& $ 4.1\,\sigma$\\ 
Luminous sources & $0\farcm5 - 4^\prime$ & 694& 8.1& 620.9& 65.0& $ 2.1\,\sigma$\\ 
Field & $0\farcm3 - 2\farcm5$& 242& 15.8& 188.5& 37.7& $ 2.4\,\sigma$\\ 
\enddata
\tablecomments{(1) Different subsets of X-ray sources. Rows 1-8 are AMUSE-Virgo galaxies, and the last row is AMUSE-Field galaxies. Luminous sources refer to $L_{\rm X} \geq 5\times 10^{38}{\rm~erg~s^{-1}}$; (2) Number of observed X-ray sources; (3) Number of LMXBs associated with the target ETGs; (4) Number of cosmic X-ray background; (5) Number of excess sources; (6) The significance of excess. 
}
\label{tab:subset}
\end{deluxetable}


\begin{thebibliography}{}
\bibitem[B\"{o}hringer et al.(1994)]{boh94} B\"{o}hringer, H., Briel, U.G., Schwarz, R.A., et al. 1994, Nature, 368, 828
\bibitem[Brandt \& Podsiadlowski(1995)]{bra95} Brandt, W.N, \& Podsiadlowski, Ph. 1995, MNRAS, 274, 461
\bibitem[Brandt, Podsiadlowski \& Sigurdsson(1995)]{bra95b} Brandt, W.N, Podsiadlowski, Ph. \& Sigurdsson, S. 1995, MNRAS, 277, L35
\bibitem[Br\"{u}ns \& Kroupa(2012)]{bru12} Br\"{u}ns, R.C., \& Kroupa, P. 2012, A\&A, 547, 65
\bibitem[Burke et al.(2015)]{bur15} Burke, C., Hilton, M., \& Collins, C. 2015, MNRAS, 449, 2353
\bibitem[Ciotti (1991)]{cio91} Ciotti, L. 1991, A\&A, 249, 99
\bibitem[Desjardins et al.(2014)]{Des14} Desjardins, T.D., Gallagher, S.C., Hornschemeier, A.E., et al. 2014, ApJ, 790, 132
\bibitem[Ebrero et al.(2009)]{2009A&A...500..749E} Ebrero, J., Mateos, S., Stewart, G.~C., Carrera, F.~J., \& Watson, M.~G.\ 2009, \aap, 500, 749
\bibitem[Fabbiano(2006)]{fab06} Fabbiano, G. 2006, ARA\&A, 44, 323
\bibitem[Ferrarese et al.(2006)]{2006ApJS..164..334F} Ferrarese, L., C{\^o}t{\'e}, P., Jord{\'a}n, A., et al.\ 2006, \apjs, 164, 334
\bibitem[Ferrarese et al.(2012)]{fer12} Ferrarese, L., C{\^o}t{\'e}, P., Cuillandre, J.-C., et al. 2012, ApJS, 200, 4
\bibitem[Ferrarese et al.(2016)]{fer16} Ferrarese, L., C{\^o}t{\'e}, P., S\'{a}chez-Janssen, R., et al. 2016, ApJ, 824, 10
\bibitem[Ferguson \& Binggeli(1994)]{fer94} Ferguson, H.C. \& Binggeli, B. 1994, Astronomy \& Astrophysics Review, 6, 67
\bibitem[Finoguenov et al.(2004)]{fin04} Finoguenov, A., Briel, U.G., Henry, J.P., et al. 2004, A\&A, 419, 47
\bibitem[Fragos et al.(2008)]{fra08} Fragos, T., Kalogera, V., Belczynski, K., et al. 2008, ApJ, 683, 346
\bibitem[Gallo et al.(2008)]{2008ApJ...680..154G} Gallo, E., Treu, T., Jacob, J., et al.\ 2008, \apj, 680, 154-168
\bibitem[Gallo et al.(2010)]{2010ApJ...714...25G} Gallo, E., Treu, T., Marshall, P.~J., et al.\ 2010, \apj, 714, 25
\bibitem[Georgakakis et al.(2008)]{2008MNRAS.388.1205G} Georgakakis, A., Nandra, K., Laird, E.~S., Aird, J., \& Trichas, M.\ 2008, \mnras, 388, 1205
\bibitem[Gilfanov(2004)]{2004MNRAS.349..146G} Gilfanov, M.\ 2004, \mnras, 349, 146
\bibitem[Gonzalez et al.(2007)]{gon07} Gonzalez, A.H., Zaritsky, D., \& Zabludoff, A.I. 2007, ApJ, 666, 147
\bibitem[Hornschemeier et al.(2006]{hor06} Hornschemeier, A.E., Mobasher, B., Alexander, D.M., et al. 2006, ApJ, 643, 144
\bibitem[Hou \& Li(2016)]{2016ApJ...819..164H} Hou, M., \& Li, Z.\ 2016, \apj, 819, 164
\bibitem[Jeltema et al.(2007)]{Jel07} Jeltema, T.E., Mulchaey, J.S., Lubin, L.M., Fassnacht, C.D. 2007, ApJ, 658, 865
\bibitem[Jeltema et al.(2008)]{Jel08} Jeltema, T.E., Binder, B., Mulchaey, J.S. 2008, ApJ, 679, 1162
\bibitem[Jord\'{a}n et al.(2005)]{jor05} Jord\'{a}n A., C{\^o}t{\'e}, P., Blakeslee J.P., et al. 2005, ApJ, 634, 1002 
\bibitem[Jord\'{a}n et al.(2009)]{jor09} Jord\'{a}n A., Peng, E.W., Blakeslee J.P., et al. 2009, ApJS, 180, 54
\bibitem[Kettula et al.(2015)]{ket15} Kettula, K., Giodini, S., van Uitert, E., et al. 2015, MNRAS, 451, 1460
\bibitem[Kim et al.(2009)]{kim09} Kim, D.-W., Fabbiano, G., Brassington, N.J., et al. 2009, ApJ, 703, 829
\bibitem[Kim et al.(2013)]{kim13} Kim, D.-W., Fabbiano, G., Ivanova, N., et al. 2013, ApJ, 764, 98
\bibitem[Kim et al.(2014)]{2014ApJS..215...22K} Kim, S., Rey, S.-C., Jerjen, H., et al.\ 2014, \apjs, 215, 22
\bibitem[Lahav \& Saslaw(1992)]{1992ApJ...396..430L} Lahav, O., \& Saslaw, W.~C.\ 1992, \apj, 396, 430
\bibitem[Lehmer et al.(2014)]{leh14} Lehmer, B.D., Berkeley, M., Zezas, A., et al. 2014, ApJ, 789, 52
\bibitem[Li et al.(2010)]{2010ApJ...721.1368L} Li, Z., Spitler, L.~R., Jones, C., et al.\ 2010, \apj, 721, 1368
\bibitem[Li et al.(2011)]{li11} Li, Z., Jones, C., Forman, W.R., et al. 2011, ApJ, 730, 84
\bibitem[Liu et al.(2015)]{liu15} Liu, C., Peng, E.W., C{\^o}t{\'e}, P., et al. 2015, ApJ, 812, 34
\bibitem[Luo et al.(2010)]{luo10} Luo, B., Brandt, W.N., Xue, Y.Q., et al. 2010, ApJS, 187, 560
\bibitem[Maccarone et al.(2007)]{mac07} Maccarone, T.J., Kundu, A., Zepf, S.E., Rhode, K.L. 2007, Nature, 445, 183
\bibitem[Martini et al.(2007)]{mar07} Martini, P., Mulchaey, J.S., Kelson, D.D. 2007, ApJ, 664, 761
\bibitem[Mei et al.(2007)]{mei07} Mei, S., Blakeslee, J.P., C\^{o}t\'{e}, P., et al. 2007, ApJ, 655, 144
\bibitem[Merloni et al.(2012)]{mer12} Merloni, A., Predehl, P., Becker, W., et al. 2012, arXiv:1209.3114
\bibitem[Merritt, Schnittman \& Komossa(2009)]{mer09} Merritt, D., Schnittman, J.D., Komossa, S. 2009, ApJ, 699, 1690
\bibitem[Mihos et al.(2005)]{mih05} Mihos, J.C., Harding, P., Feldmeier, J., Morrison, H. 2005, ApJ, 631, L41
\bibitem[Mihos(2015)]{mih15} Mihos, J.C., 2015, in {\it The General Assembly of Galaxy Halos: Structure, Origin and Evolution}, Proceedings of the International Astronomical Union, IAU Symposium, Volume 317, pp. 27-34
\bibitem[Mihos et al.(2017)]{mih17} Mihos, J.C., Harding, P., Feldmeier, J., et al. 2017, ApJ, 834, 16
\bibitem[Miller et al.(2012a)]{2012ApJ...747...57M} Miller, B., Gallo, E., Treu, T., \& Woo, J.-H.\ 2012a, ApJ, 745, L13
\bibitem[Miller et al.(2012b)]{2012ApJ...747...57M} Miller, B., Gallo, E., Treu, T., \& Woo, J.-H.\ 2012b, \apj, 747, 57
\bibitem[Mineo et al.(2014)]{min14} Mineo, S., Fabbiano, G., D'Abrusco, R., et al. 2014, ApJ, 780, 132
\bibitem[Morishita et al.(2016)]{mor16} Morishita, T., Abramson L.E., Treu, T., et al. 2016, ApJL, submitted (arXiv:1610.08503)
\bibitem[Pandya et al.(2016)]{pan16} Pandya, V., Mulchaey, J., \& Greene, J.E. 2016, ApJ, 819, 162
\bibitem[Park et al.(2006)]{2006ApJ...652..610P} Park, T., Kashyap, V.~L., Siemiginowska, A., et al.\ 2006, \apj, 652, 610
\bibitem[Peacock \& Zepf(2016)]{pea16} Peacock, M.B., \& Zepf, S.E., 2016, ApJ, 818, 33
\bibitem[Phillipps et al.(2001)]{phi01} Phillipps, S., Drinkwater, M.J., Gregg, M.D., \& Jones, J.B. 2001, ApJ, 560, 201
\bibitem[Sarazin et al.(2000)]{2000ApJ...544L.101S} Sarazin, C.~L., Irwin, J.~A., \& Bregman, J.~N.\ 2000, \apjl, 544, L101
\bibitem[Sazonov et al.(2006)]{saz06} Sazonov, S., Revnivtsev, M., Gilfanov, M., Churazov, E. 2006, A\&A, 450, 117
\bibitem[Seth et al.(2014)]{set14} Seth, A.C., van den Bosch, R., Mieske, S., et al. 2014, Nature, 513, 398
\bibitem[Schnittman(2013)]{sch13} Schnittman, J.D. 2013, Classical and Quantum Gravity, 30, 244007
\bibitem[Shen et al.(2007)]{Shen07} Shen, Y., Mulchaey, J.S., Raychaudhury, S., et al. 2007, ApJ, 654, L115
\bibitem[Strader et al.(2013)]{str13} Strader, J., Seth, A.C., Forbes, D.A., et al. 2013, ApJ, 775, L6
\bibitem[Trinchieri \& Fabbiano(1985)]{tri85} Trinchieri, G., \& Fabbiano, G. 1985, ApJ, 296, 447
\bibitem[Totsuji \& Kihara(1969)]{1969PASJ...21..221T} Totsuji, H., \& Kihara, T.\ 1969, \pasj, 21, 221
\bibitem[Tzanavaris et al.(2014)]{tza14} Tzanavaris, P., Gallagher, S.C., Hornschemeier, A.E., et al. 2014, ApJS, 212, 9
\bibitem[Tzanavaris et al.(2016)]{tza16} Tzanavaris, P., Hornschemeier, A.E., Gallagher, S.C., 2016, ApJ, 817, 95
\bibitem[Urban et al.(2011)]{urb11} Urban, O., Werner, N., Simionescu, A., et al. 2011, MNRAS, 414, 2101
\bibitem[Wang(2004)]{2004ApJ...612..159W} Wang, Q.~D.\ 2004, \apj, 612, 159
\bibitem[Wu(2001)]{wu01} Wu, K. 2001, PASA, 18, 443
\bibitem[Zhang et al.(2011)]{2011A&A...533A..33Z} Zhang, Z., Gilfanov, M., Voss, R., et al.\ 2011, \aap, 533, 33
\bibitem[Zhang et al.(2013)]{2013A&A...556A...9Z} Zhang, Z., Gilfanov, M., \& Bogd{\'a}n, {\'A}.\ 2013, \aap, 556, 9
\bibitem[Zibetti et al.(2005)]{zib05} Zibetti, S., White, S.D.M., Schneider, D.P., Brinkmann, J. 2005, MNRAS, 358, 949
\bibitem[Zuo et al.(2008)]{zuo08} Zuo, Z.-Y., Li, X.-D., \& Liu, X.-W. 2008, MNRAS, 387, 121
\end{thebibliography}
\end{document}